\documentclass[11pt]{article}   	
\usepackage{geometry}                		
\usepackage{graphicx}				

\usepackage{amssymb}
\usepackage{amsmath}
\usepackage{authblk}
\usepackage[table]{xcolor}
\usepackage{multirow}
\usepackage{graphicx}
\usepackage[normalem]{ulem}

\newcommand{\be}{\begin{equation}}
\newcommand{\ee}{\end{equation}}
\newcommand{\pom}{{I\!\!P}}
\newcommand{\regg}{{I\!\!R}}

\usepackage[utf8]{inputenc}
\usepackage{xspace}
\usepackage{lineno} 

\usepackage[hypertexnames,
    pdftex,%
    colorlinks,%
		citecolor=blue,%
    hyperindex,%
    plainpages=false,%
    bookmarksopen,%
    bookmarksnumbered%
  ]{hyperref}
\usepackage[numbers,sort&compress]{natbib}
\newcommand\Eq[1]{(\ref{#1})}
\newcommand\Fig[1]{Fig.~\ref{#1}}

\DeclareRobustCommand\GeV{\ensuremath{\mathrm{GeV}}\xspace}

\DeclareRobustCommand\estat{\ensuremath{\delta_\mathrm{stat}}\xspace}

\DeclareRobustCommand\sred{\ensuremath{\sigma_\mathrm{red}}\xspace}

\DeclareRobustCommand\srDD{\ensuremath{\sigma_\mathrm{red}^\mathrm{D(4)}}\xspace}

\DeclareRobustCommand\chiNDF{\ensuremath{\chi^2/\mathrm{NDF}}\xspace}

\DeclareRobustCommand\ifb{\ensuremath{\mathrm{fb}^{-1}}\xspace}

\newcommand\DD{\mathrm{D}}

\definecolor{eM}{hsb}{0.3,0.5,1}
\definecolor{eS}{hsb}{0,1,0.6}

\usepackage{color}

\title{\bf Extracting the partonic structure of colorless exchanges at the Electron Ion Collider}
\author[1]{N\'estor Armesto}
\author[2]{Paul R. Newman}
\author[3]{Wojciech S\l{}omi\'nski}
\author[4]{ Anna M. Sta\'sto}
\affil[1]{\small \it Instituto Galego de F\'{\i}sica de Altas Enerx\'{\i}as IGFAE,
Universidade de Santiago de Compostela, 15782 Santiago de Compostela, Galicia-Spain}
\affil[2]{\small \it School of Physics and Astronomy, University of Birmingham, B15 2TT, UK}
\affil[3]{\small \it Institute of Theoretical Physics, Jagiellonian University, Kraków, Poland}
\affil[4]{\small \it Department of Physics, Penn State University, University Park, PA 16802, USA}

\begin{document}
\maketitle
\begin{abstract}
 We investigate the 
determination of the
partonic structure of colorless exchanges
in deep inelastic diffractive $ep$ scattering
at the Electron Ion Collider,
using the standard decomposition into Pomeron and Reggeon
contributions.
We perform fits to simulated 
diffractive cross section
pseudodata 
in four variables, including the momentum transfer $t$,
to estimate the achievable precision    
on the Pomeron and Reggeon quark and gluon distributions.
We analyze the influence of different cuts in the 
kinematic variables, beam energy configurations and luminosities,
including a `first year' scenario. 
We conclude that the EIC will be able to constrain the partonic structure of the sub-leading Reggeon exchange with a  
precision comparable to that of the leading Pomeron exchange.
\end{abstract}

\section{Introduction}
\label{sec:intro}

Processes with rapidity gaps, or `diffractive' processes, have been extensively studied in Deep Inelastic electron-proton Scattering (DIS) at the HERA collider at DESY~\cite{Adloff:1997sc,Breitweg:1997aa}, see the review~\cite{Newman:2013ada} and references therein. These events  are characterized by the presence of a wide rapidity region 
that does not contain any activity, 
adjacent to the final state proton which scatters elastically or dissociates into small mass excitations. The interpretation of diffraction requires that it be mediated by a colorless exchange, usually referred to as the Pomeron 
in the high energy limit.
When the  
modulus of the photon virtuality, $Q^2$, is high enough, it provides a hard scale, allowing perturbative Quantum Chromodynamics (QCD) to be applied.
In close analogy to the inclusive DIS case, the diffractive cross section can then be computed using a collinear factorization theorem~\cite{Collins:1997sr}, where hard scattering coefficients are combined  with the diffractive parton distribution functions (DPDFs) of the proton \cite{Berera:1995fj}.  For the description of the diffractive data at HERA, it was shown that 
the DPDFs could be 
interpreted as the parton distributions 
of the Pomeron convoluted with a flux factor
that could be expressed in terms of an associated Regge trajectory,
dependent on 
the momentum transfer $t$ and the longitudinal momentum fraction $\xi$ of the proton carried by the colorless exchange \cite{Ingelman:1984ns}. 
This paradigm worked very well for the description of the data at small values of $\xi$. However, in order to describe the data at 
the largest values of $\xi$, it was necessary to include a secondary sub-leading 
`Reggeon' exchange, 
consistent with the approximately degenerate 
Regge trajectory describing the $\rho,\omega,a_2$ and $f_2$ mesons.  
The parton densities of the Pomeron were well
constrained by the inclusive diffractive HERA data
from the H1 and ZEUS collaborations, with additional constraints on the large 
momentum fraction
behavior of the gluon coming from  diffractive dijet 
data~\cite{Chekanov:2005vv, Aktas:2006hx, Aktas:2006hy,Chekanov:2008fh,Chekanov:2009aa,Aaron:2010aa,Aaron:2012ad}. On the other hand, 
in the kinematic region of the measurements, it was not possible
to extract the partonic content of the secondary Reggeon with
any degree of accuracy. Instead, it was typically taken
from a parametrization derived from fits to pion structure function data~\cite{Gluck:1991ey}.

Diffractive processes will be studied with high precision at the Electron Ion Collider (EIC)~\cite{Accardi:2012qut,AbdulKhalek:2021gbh}, 
a facility that will be built
at Brookhaven National Laboratory (BNL) using and
upgrading the existing accelerator complex for the Relativistic Heavy Ion Collider (RHIC).
The machine will operate with high instantaneous luminosity, of order $10^{33}-10^{34} \; {\rm cm}^{-2} {\rm s}^{-1}$, and  at variable center of mass energy $\sqrt{s}=20-140 \, \rm GeV$. It will be able to collide electrons with protons and a wide range of nuclei, and will have the possibility of various combinations of polarization  both for electrons and protons as well as for light ions. 
With dedicated forward instrumentation included as a fundamental
design consideration from the outset, EIC experiments will have 
very good  acceptance for elastically scattered protons in a wide range of 
$t$ and $\xi$. 
The wide scattered proton acceptance, together with the
potential to vary the beam energies
in a large range spanning the region between fixed target facilities and 
HERA,
gives the EIC great potential to explore the nature of both 
the Pomeron and the secondary Reggeon, and to map the transition between 
them 
in diffractive DIS. 

In this paper we perform a detailed study of the possibilities of the EIC for extracting the partonic content of 
colorless exchanges, with the emphasis on the Reggeon. 
We simulate pseudodata for the four-dimensional diffractive cross section, including the momentum transfer $t$ dependence, 
which can be measured with percent accuracy up to about $-t \sim 1.5 \; \rm GeV^2$ because of the high projected luminosity. 
Using these pseudodata we perform a series of fits with Pomeron and Reggeon contributions
evolving with $\beta = x /\xi$ and $Q^2$ according to the 
Dokshitzer-Gribov-Lipatov-Altarelli-Parisi (DGLAP) equations~\cite{Gribov:1972ri,Gribov:1972rt,Dokshitzer:1977sg,Altarelli:1977zs}. 
This is the first time that 
the Reggeon component 
has been treated on an equal footing with the Pomeron contribution
in such a fit. 

The structure of the paper is as follows. In the next section we present the basic definitions of the variables and expressions for the cross sections used in  diffractive DIS. We also  discuss the experimental setup of the EIC in the context of diffractive measurements. In Sec.~\ref{sec:setup}  we present the method for generating the EIC pseudodata and discuss the details of the fits of Pomeron and Reggeon parton distributions. In Sec.~\ref{sec:results} we show and discuss the results of these fits, for two beam energy scenarios.
Finally in Sec.~\ref{sec:conclu}, we state our conclusions and present an outlook.

\section{Definitions and kinematics}
\label{sec:kinematics}

\subsection{Diffractive variables and definitions}
\label{subsec:variables}

\begin{figure}[htb]
\centerline{
\includegraphics[width=0.7\columnwidth, clip]{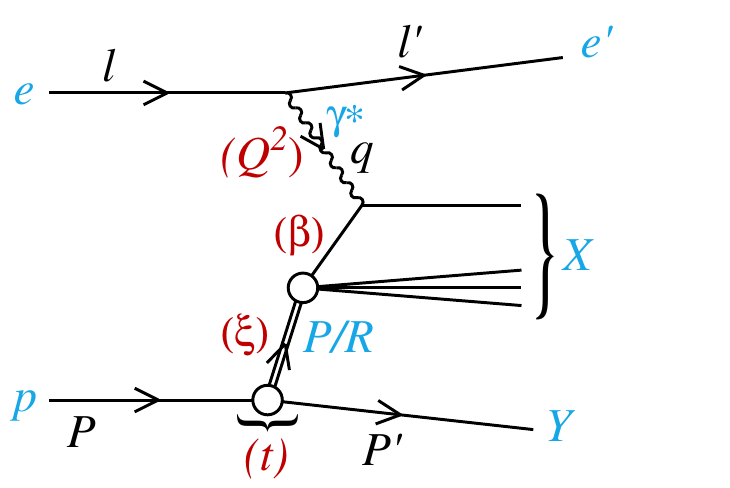}
}
\caption{Diagram showing the neutral current diffractive DIS process and the relevant kinematic variables in the one photon exchange approximation.}
\label{fig:kinvar}
\end{figure}

In this work we focus on neutral current diffractive Deep Inelastic Scattering (DDIS) in the one photon exchange approximation. We do not account for radiative corrections, but in principle these  can be included. We do not consider the possibility of charged currents or $Z^0$ exchange, which at EIC energies give a very small contribution for the purposes of this study.
For an electron or positron with four momentum $l$ and a proton with four-momentum $P$, the corresponding diagram is shown in Fig.~\ref{fig:kinvar}.
A characteristic feature of the diffractive process, as illustrated in Fig.~\ref{fig:kinvar}, is the presence of
a very forward final proton  
separated by a rapidity gap  from  the 
remainder of the hadronic final state (system $X$). It is mediated by the colorless object, indicated by $P/R$, to which we refer generally as `diffractive exchange'.

The usual DDIS variables, defined
in terms of the four-momenta indicated in 
Fig.~\ref{fig:kinvar}, are
\begin{eqnarray}
\nonumber \\
Q^2 &=& -q^2\ , \nonumber \\
y &=& \frac{P\cdot q}{P\cdot \ell}\ , \nonumber \\
x &=& \frac {Q^2}{2 P\cdot q}
    = \frac{Q^2}{y s}
    \ ,\nonumber \\
\beta &=& \frac{Q^2}{2\, (P-P^\prime) \cdot q}\ ,
\nonumber \\
\xi &=& \frac{x}{\beta}\ ,\nonumber \\
t&=&(P^\prime-P)^2\ .
\end{eqnarray}
Among them are the standard DIS variables: $Q^2$ the (negative) photon virtuality, $s$ the square of the total center-of-mass energy, Bjorken $x$ and the inelasticity $y$.
 In diffractive DIS (DDIS), additional variables are introduced:
  $t$ is the squared
 four-momentum transfer   at the proton vertex, $\xi$ (alternatively denoted by $x_\pom$
 in the HERA literature)  can be interpreted as  the momentum fraction of the diffractive exchange   with respect to the beam hadron, and $\beta$ 
is the momentum fraction of the quark coupling to the virtual photon with respect to the diffractive exchange. 
The variables $\xi$ and $\beta$ are related to the Bjorken $x$ variable by $x=\xi\beta$.
The variable $z$ is commonly introduced, generalizing $\beta$ to
the momentum fraction of any parton (i.e., also including gluons) relative
to the diffractive exchange.

\begin{figure}[htb]
\centerline{
\includegraphics[width=1.0\columnwidth, clip,trim=0 6 0 0]{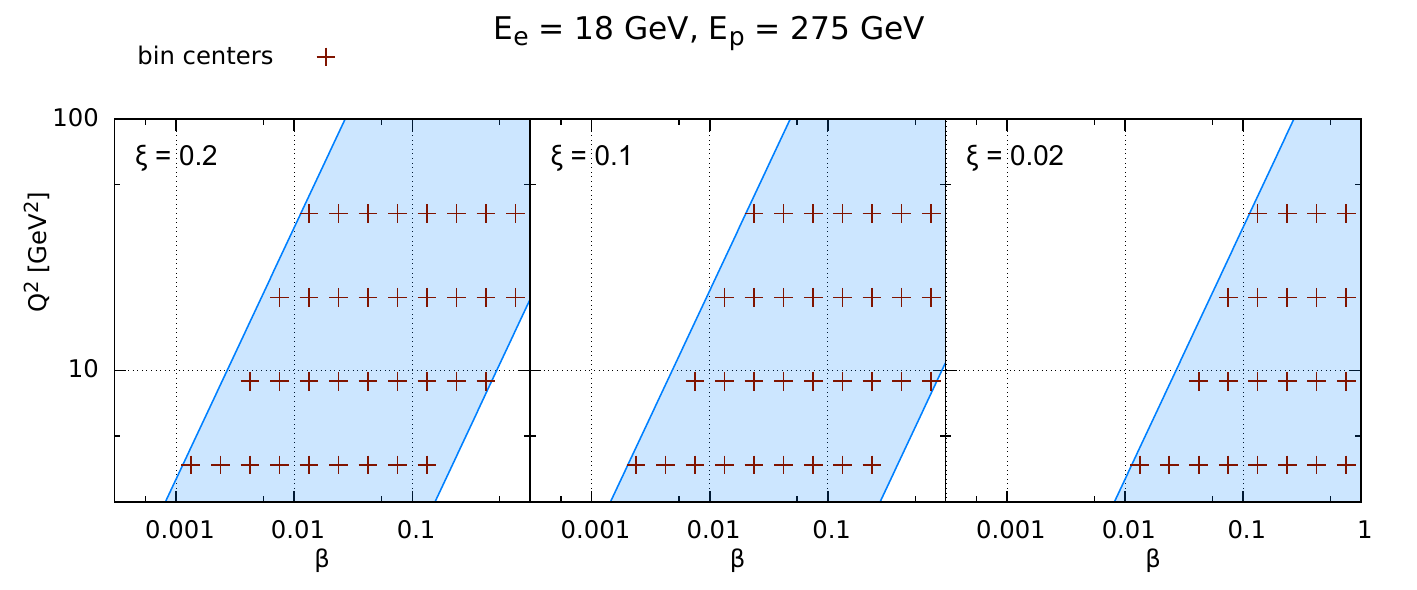}
}
\centerline{
\includegraphics[width=1.0\columnwidth, clip,trim=0 0 0 -10]{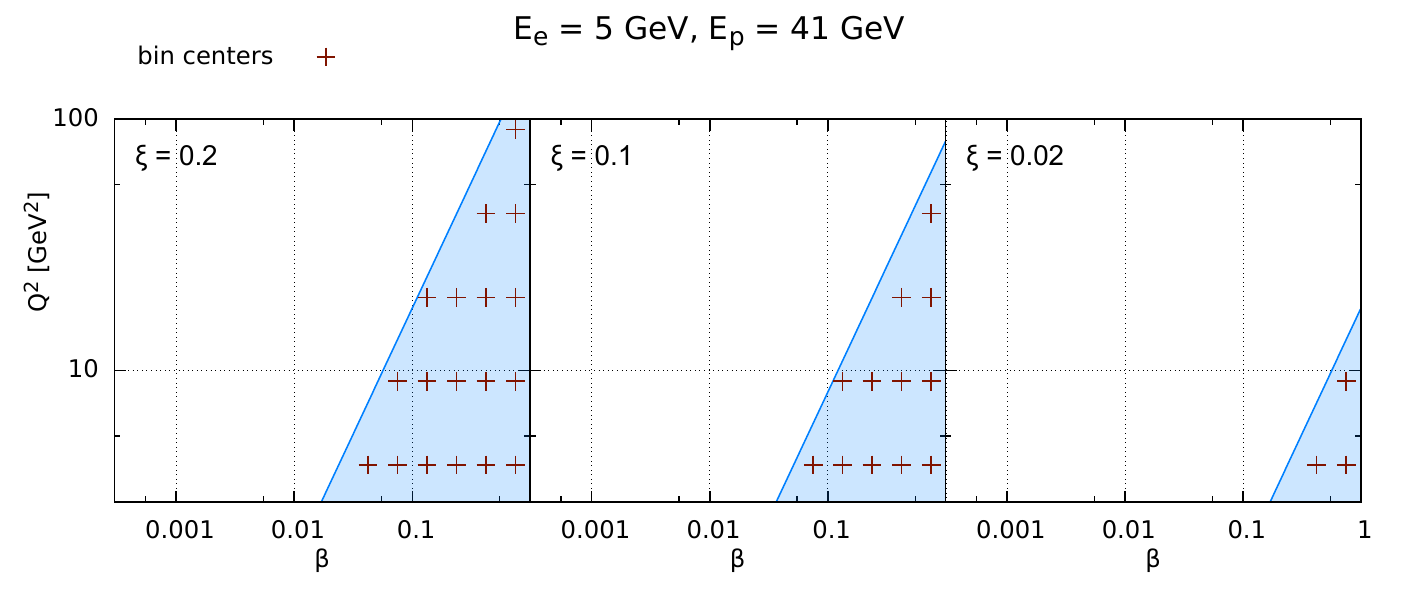}
}
 \caption{Kinematic $(\beta,Q^2)$ range for selected \(\xi\) bins, covered by the EIC and used in our simulations.
 The upper and lower plots correspond to the two $E_p\times E_e$ energy setups: 275 GeV $\times$ 18 GeV and 41 GeV $\times$ 5 GeV, respectively.
 The crosses mark the bin centers of the simulated data.}
\label{fig:kin}
\end{figure}

In Fig.~\ref{fig:kin} we show the kinematic coverage in $\beta$ and $Q^2$ for fixed $\xi$ of the EIC for two   $E_p\times E_e$ beam energy setups: 275 GeV $\times$ 18 GeV and 41 GeV $\times$ 5 GeV.  
We note that 
with the currently planned instrumentation around the outgoing 
proton beamline (Section~\ref{subsec:ptag}), the EIC will be able to cover the 
$\xi >0.1$ region, which was inaccessible for HERA. 

The measured cross section is differential in four variables, usually chosen to be $\beta,\xi,Q^2$ and $t$:
\be
\frac{d^4 \sigma^{\DD}}{d\xi d\beta dQ^2 dt} = \frac{2\pi \alpha_{\rm em}^2}{\beta Q^4} \, Y_+ \, \sred^{\DD(4)}\,  ,
\label{eq:sigmared4}
\ee
where we introduced the reduced cross section $\sred^{\DD(4)}$ and defined 
$Y_+= 1+(1-y)^2$.
Upon integration over the momentum transfer $t$ one can define 
\be
\label{eq:sigmared3}
\frac{d^3 \sigma^{\DD}}{d\xi d\beta dQ^2} = \frac{2\pi \alpha_{\rm em}^2}{\beta Q^4} \, Y_+ \, \sred^{\DD(3)}\, .
\ee

In the one photon exchange approximation, the reduced cross sections can be expressed in terms of two diffractive structure functions
$F_2^{\DD}$ and $F_\mathrm{L}^{\DD}$:
\be
\sred^{\DD(4)} = F_2^{\DD(4)}(\beta,\xi,Q^2,t) - \frac{y^2}{Y_+} F_\mathrm{L}^{\DD(4)}(\beta,\xi,Q^2,t) \; ,
\label{eq:sred4}
\ee
\be
\label{eq:sred3}
\sred^{\DD(3)} = F_2^{\DD(3)}(\beta,\xi,Q^2) - \frac{y^2}{Y_+} F_\mathrm{L}^{\DD(3)}(\beta,\xi,Q^2) \; .
\ee
Note that the dimensions of these structure functions depend on the number of variables, i.e.,  $F_\mathrm{2,L}^{\DD(4)}$ have dimension $\GeV^{-2}$ and $F_\mathrm{2,L}^{\DD(3)}$ are dimensionless.

The reduced cross sections $\sred^{\DD(3)}$ and (to some extent) $\sred^{\DD(4)}$ have been measured at HERA~\cite{Adloff:1997sc,Breitweg:1997aa,Chekanov:2005vv, Aktas:2006hx, Aktas:2006hy,Chekanov:2008fh,Chekanov:2009aa,Aaron:2010aa,Aaron:2012ad}. These data have been used as inputs to perturbative QCD analyses based on collinear factorization~\cite{Collins:1997sr,Berera:1995fj,Trentadue:1993ka}, where
the diffractive cross section reads
\be
d\sigma^{ep\rightarrow eXY}(\beta,\xi,Q^2,t) \; = \; \sum_i \int_{\beta}^{1} dz \ d\hat{\sigma}^{ei}\left(\frac{\beta}{z},Q^2\right) \, f_i^{\rm D}(z,\xi,Q^2,t) \; ,
\label{eq:collfac}
\ee
up to terms of  order  ${\cal O}(1/Q^2)$. Here,
the sum is performed over all parton species (gluon and all quark flavors).
The hard scattering partonic cross section $d\hat{\sigma}^{ei}$ can be computed perturbatively in QCD and is the same as in  inclusive deep inelastic scattering. The long distance 
parts $f_i^{\rm D}$ are the  diffractive parton distribution functions (DPDFs),
which can be interpreted as conditional probabilities for partons in the proton, provided the proton is scattered into the final state system $Y$ with  four-momentum $P^\prime$. 
They are non-perturbative objects to be extracted from data, but their evolution through the DGLAP evolution equations~\cite{Gribov:1972rt,Gribov:1972ri,Altarelli:1977zs,Dokshitzer:1977sg} can be computed in perturbative QCD,
similarly to the case of standard PDFs in  inclusive DIS.

\subsection{Experimental Considerations}
\label{subsec:ptag}

The EIC project has a well-established plan for an experimental programme
with a detector design specified from
physics measurement target considerations in~\cite{AbdulKhalek:2021gbh},
evolved through proto-collaborations~\cite{Adkins:2022jfp,ATHENA:2022hxb} and culminating in the current
design being developed by the ePIC collaboration~\cite{epic} 
for location at IP6 of the Brookhaven 
RHIC collider. Plans for a second,
complementary, detector at IP8 are also under development. 
The work presented here is not specific to any particular experimental 
design, though it does assume comprehensive beam-line instrumentation, 
as anticipated in all proposals mentioned above. 
Whilst the acceptance range is taken to be extended considerably compared with
HERA, we have only assumed a modest evolution in performance,
leading to systematic uncertainties that are sufficient for the current work, 
but are less aggressive than the actual experimental targets. 

Since the current work is strongly dependent on wide acceptance in
the scattered proton-related variables $\xi$ and $t$, we have assumed the
use of the full range of scattered proton detectors planned for ePIC, 
using both Roman pot near-beam technologies and also `B0' detectors 
surrounding the beam-pipe, giving near-complete 
acceptance\footnote{The small gaps in the angular acceptance,
as shown for example in Fig. 4 of \cite{Armesto:2021fws}, lead to 
corresponding
acceptance limitations 
in $t$, which are neglected here.} in the 
proton scattering angular range
$1 < \theta < 20 \ {\rm mrad}$. 
As discussed in more detail in~\cite{Armesto:2021fws} (see in particular Figure 4),
the corresponding acceptance extends over a wide $\xi$ range,
reaching beyond the point where exchange particle models become unreliable. 
The acceptance in $|t|$ varies with the proton beam energy and $\xi$, 
extending from the kinematic limit to several ${\rm GeV}^2$ at 
$E_p = 275 \ {\rm GeV}$ and $\xi \rightarrow 1$, but covering only approximately 
$0.1 < |t| < 0.6 \ {\rm GeV^2}$ at $E_p = 41 \ {\rm GeV}$ and $\xi = 0.3$. 
The resolutions in $\xi$ and $t$ in the EIC forward detectors are expected to
be comfortably better than those at HERA, and calibrations relative to variables 
reconstructed in the central detector -- e.g., produced particle masses 
and transverse momenta in exclusive vector meson processes -- will prevent significant 
biases. However, since the details of the expected performance are hard to
quantify at this stage, we have made two alternative assumptions, referred
to here as the `dense' and `sparse' binnings. 
The resolutions achievable on $x$ and $Q^2$ in 
the central EIC detectors and already 
considered in simulations of inclusive DIS 
measurements (see, e.g.,~\cite{Cerci:2023uhu,Armesto:2023hnw})
are applicable to $\beta$ and $Q^2$ reconstruction in the diffractive case and 
are comfortably sufficient for the binning assumptions made here. 

Systematic uncertainties on the measurements 
that are uncorrelated between data points are conservatively taken to be at the
5\% level, a precision similar to that 
already achieved in proton-tagged measurements at HERA~\cite{H1:2012xlc} and
less demanding than assumed in our previous work~\cite{Armesto:2021fws}. 
In extracting DPDFs, normalization uncertainties, for example due to the
luminosity measurement or Roman pot proton reconstruction efficiencies, 
are potentially more important, since they 
propagate into normalization uncertainties on the DPDFs themselves.
Here, we assume 2\%, similarly to~\cite{Armesto:2021fws}.

Several beam energy configurations are proposed at the EIC. Here, we 
simulate samples at the lowest and highest planned center of mass
energies ($29 \ {\rm GeV}$ and $140 \ {\rm GeV}$, respectively), with 
integrated luminosities
corresponding to approximately one year's data taking at each energy,
as described in Section~\ref{subsec:pseudodata}. We also consider a larger integrated luminosity at the highest energy.
As was observed for the case of photoproduction in comparing HERA with 
fixed target data~\cite{H1:1997vke}, the wide lever-arms in both 
center-of-mass energy and $\xi$ lead to simultaneous strong sensitivity to both the
Pomeron and sub-leading ($1/s$-suppressed) exchanges.

\section{Method}
\label{sec:setup}

In this Section we explain the method for generating the EIC pseudodata, which follows that employed in~\cite{Armesto:2021fws} where a detailed study of the accessible kinematic region at the EIC can be found. Next, we indicate how the fits of Pomeron ($\pom$) and Reggeon ($\regg$) parton distributions are done, and how their uncertainties are evaluated.

\subsection{Pseudodata generation}
\label{subsec:pseudodata}

Pseudodata are generated in all four variables $(\beta,\xi,Q^2,t)$ using the factorized expression~\eqref{eq:collfac} and the Regge factorization assumption~\cite{Ingelman:1984ns}. Specifically, we assume that the diffractive parton distributions take the form
\begin{eqnarray}
f_k^{D(4)}(z,Q^2,\xi,t) &=&
\phi_\pom(\xi,t)\, f^\pom_k(z,Q^2)
+
\phi_\regg(\xi,t)\, f^\regg_k(z,Q^2) \, ,
\label{eq:Reggefac}
\end{eqnarray}
where
\be
\phi_\mathbb{M}(\xi,t) =
\frac{e^{B_\mathbb{M} t}}{\xi^{2\alpha_\mathbb{M}(t)-1}} \, ,
\label{eq:fluxes}
\ee
and ${\alpha_\mathbb{M}(t)=\alpha_\mathbb{M}(0)+\alpha_\mathbb{M}'\,t}$ ($\mathbb{M} = \pom,\regg$)
are the Pomeron and Reggeon trajectories, taken to be linear.

Plugging this expression into~\eqref{eq:sred4} yields
\begin{eqnarray}
\srDD &=&
\phi_\pom(\xi,t)\, \mathcal{F}_2^\pom(\beta,Q^2) + \phi_\regg(\xi,t)\, \mathcal{F}_2^\regg(\beta,Q^2)
\nonumber \\
	&&- \frac{y^2}{Y_+} \left[
\phi_\pom(\xi,t)\, \mathcal{F}_L^\pom(\beta,Q^2) + \phi_\regg(\xi,t)\, \mathcal{F}_L^\regg(\beta,Q^2)
\right].
\label{eq:sred4Rfac}
\end{eqnarray}

Note that the
longitudinal contributions are multiplied by
$y^2/Y_+$, which breaks factorization by coupling $\xi$ to $(\beta,Q^2)$ at fixed $y$ and $s$.
As $\mathcal{F}_L$ are of higher QCD order, this factorization breaking is small -- especially at low $y$.

We use the parameters 
given in~\cite{Chekanov:2009aa}, namely for the fluxes:
$\alpha_\pom(0) = 1.1\,, 
\alpha^\prime_\pom = 0\,,
\alpha_\regg(0) = 0.7\,,
\alpha^\prime_\regg = 0.9\,\GeV^{-2}\,.$
For the Pomeron parton densities at the initial scale $\mu_0$, we take
\be
f^\pom_k(z, \mu_0^2) = \hat A_k\, z^{\hat B_k}\, (1-z)^{\hat C_k},
\ee
where
$\hat A_g = 0.301, \hat B_g = -0.161, \hat C_g = -0.232,
\hat A_q = 0.151, \hat B_q = 1.23, \hat C_q = -0.332$.

For the Reggeon parton densities, we also follow the commonly
adopted HERA approach, whereby
\be
f^\regg_k(z,Q^2) = A_\regg\, f^{\pi^0}_k(z,Q^2),
\label{eq:grvpdfs}
\ee
where the GRV parametrization for the pion parton densities~\cite{Gluck:1991ey} is used
with $A_\regg = 2.7$.

We consider two $E_p\times E_e$ beam energy setups: 275 GeV $\times$ 18 GeV (`high energy')  and 41 GeV $\times$ 5 GeV  (`low energy'). For the former, two integrated luminosity scenarios of 100 and 10 fb$^{-1}$ 
and two, \textsl{dense} and \textsl{sparse} binnings
are considered, while for the latter we take only 10 fb$^{-1}$ and the \textsl{sparse} binning.

The logarithmically spaced \textsl{dense} binning can be summarized as: 
\[\begin{array}{rll}
-t    \;\in & [0.01, 2] \,\GeV^2     & (23 \; \mathrm{bins});\\
\xi   \;\in & [0.0004, 0.4] & (24 \; \mathrm{bins});\\
\beta \;\in & [0.001, 1]  & (12 \; \mathrm{bins});\\
Q^2   \;\in & [3, 62] \,\GeV^2    & (4 \; \mathrm{bins}). \\
\end{array}\]

The logarithmically spaced \textsl{sparse} binning follows:
\[\begin{array}{rll}
-t    \;\in & [0.01, 2] \,\GeV^2     & (14 \; \mathrm{bins});\\
\xi   \;\in & [0.0004, 0.4] & (18 \; \mathrm{bins});\\
\beta \;\in & [0.001, 1]  & (12 \; \mathrm{bins});\\
Q^2   \;\in & [3, 130] \,\GeV^2    & (5\; \mathrm{bins}). \\
\end{array}\]

In both cases, only bins whose centers fulfill the kinematic constraint $Q^2 = \xi \beta\, y s$, with $y \in [0.005, 0.96]$ 
are considered for the fits, and it is further required that there be
at least 8 events per bin ($\delta_\mathrm{stat} < 35\%$).
For the higher energy setup 
and higher luminosity
we get 9652 and 5267 pseudodata points for the \textsl{dense} and \textsl{sparse} binnings, respectively.
For the lower luminosity the corresponding numbers are 
9482 and 5180.
For the lower energy setup, lower luminosity and \textsl{sparse} binning, we arrive at 1037 pseudodata points. Note that, when we apply additional cuts, as done in the next Section, these numbers are reduced.

Pseudodata are generated according to~\eqref{eq:sred4Rfac} for the central values, smeared with a Gaussian distribution around the central value with standard deviation given by the addition in quadrature of statistical and uncorrelated systematic uncertainties
of 5\%. In addition, we consider a normalization uncertainty of 2\%, 
which is discussed separately. 
In \Fig{fig:sigred4}
we show examples of the generated pseudodata as a function of (negative) momentum transfer $-t$ (left plot) and longitudinal momentum fraction  $\xi$ (right plot). The change in the slope in the $-t$ dependence for different values of $\xi$ is due to the 
transition from Pomeron to Reggeon dominated regions. Similarly, the change in the behavior of the cross section as a function of $\xi$ at around $0.07-0.08$ corresponds to the boundary between the regions where the two exchanges are dominant. 
The relative contributions of the two components can also be visualized by plotting the Reggeon to Pomeron ratio in the cross section, which is shown 
as a function of $-t$ for fixed values of $\xi$ and $\beta$  in 
Fig.~\ref{fig:ratioR2P}.
The strongest dependence is on $\xi$, with the ratio changing from $<1$ for small $\xi \sim 0.02$ to $>1$ for $\xi \sim 0.1$ or greater. 
The ratio increases with increasing $-t$ for large values of $\xi$ and decreases with increasing $-t$ for small $\xi$, reflecting the different slopes assumed in the parametrization.
The region $\xi>0.1$, where the Reggeon is dominant, was not accessible at HERA. 
For  $|t| \lesssim 1.2$ GeV$^2$, less than 4\% of data points have $\estat >5$\% for the integrated luminosity of 10~\ifb, and all data points have $\estat<4.3$\% for the integrated luminosity of 100~\ifb.

\begin{figure}[htb]
\centerline{
\includegraphics[width=0.49\columnwidth,clip,trim=0 5 0 0]{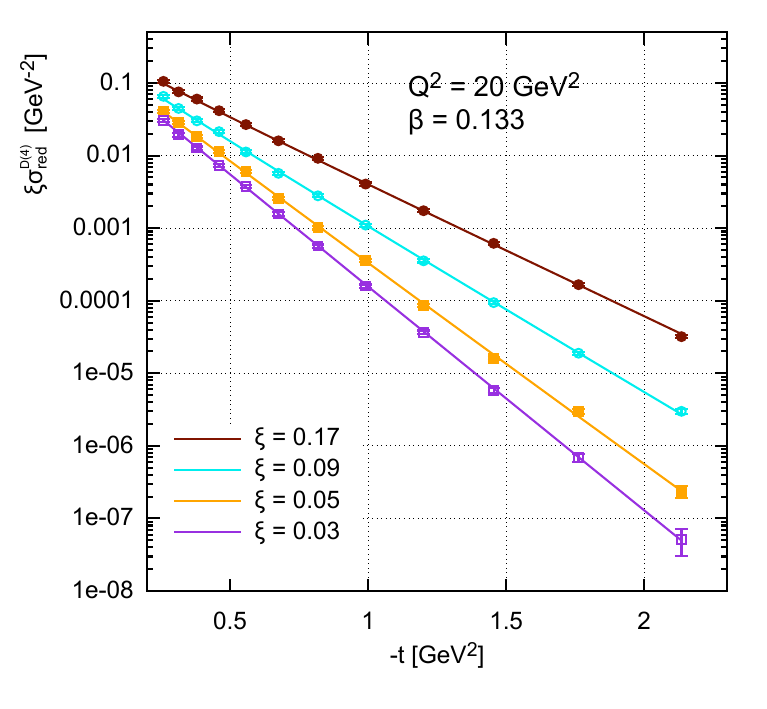}
\includegraphics[width=0.49\columnwidth,clip,trim=0 0 0 0]{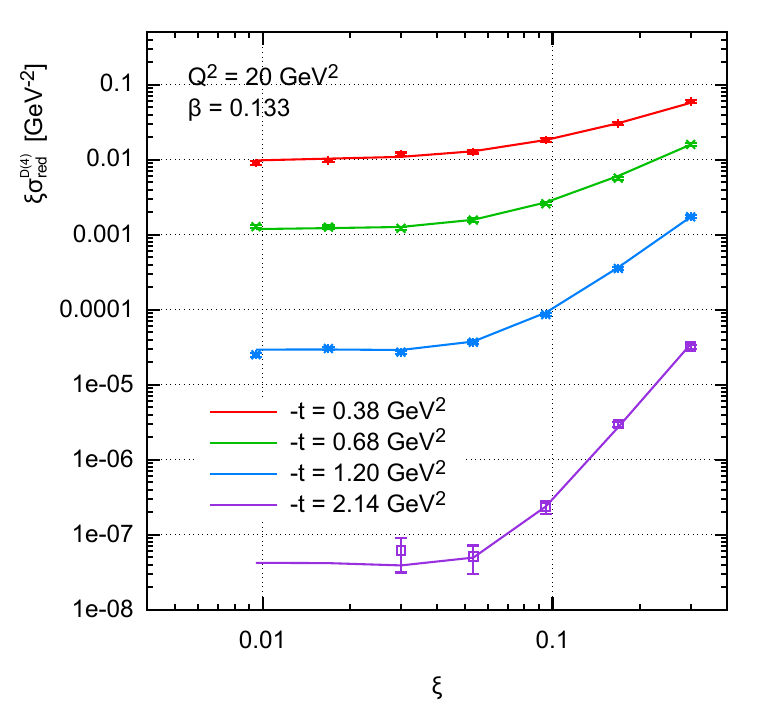}
}
\caption{Reduced cross section versus $-t$ (left) and $\xi$ (right) for \(Q^2 = 20\,\GeV^2,\, \beta=0.133\) and selected values of $\xi,t$, for the top energy setup and integrated luminosity of 100 fb$^{-1}$. Pseudodata, given by the points with error bars, are compared to the predictions of the model (lines) providing the central values.}
\label{fig:sigred4}
\end{figure}

\begin{figure}[htb]
\centerline{
\includegraphics[width=0.99\columnwidth,clip,trim=0 15 0 45]{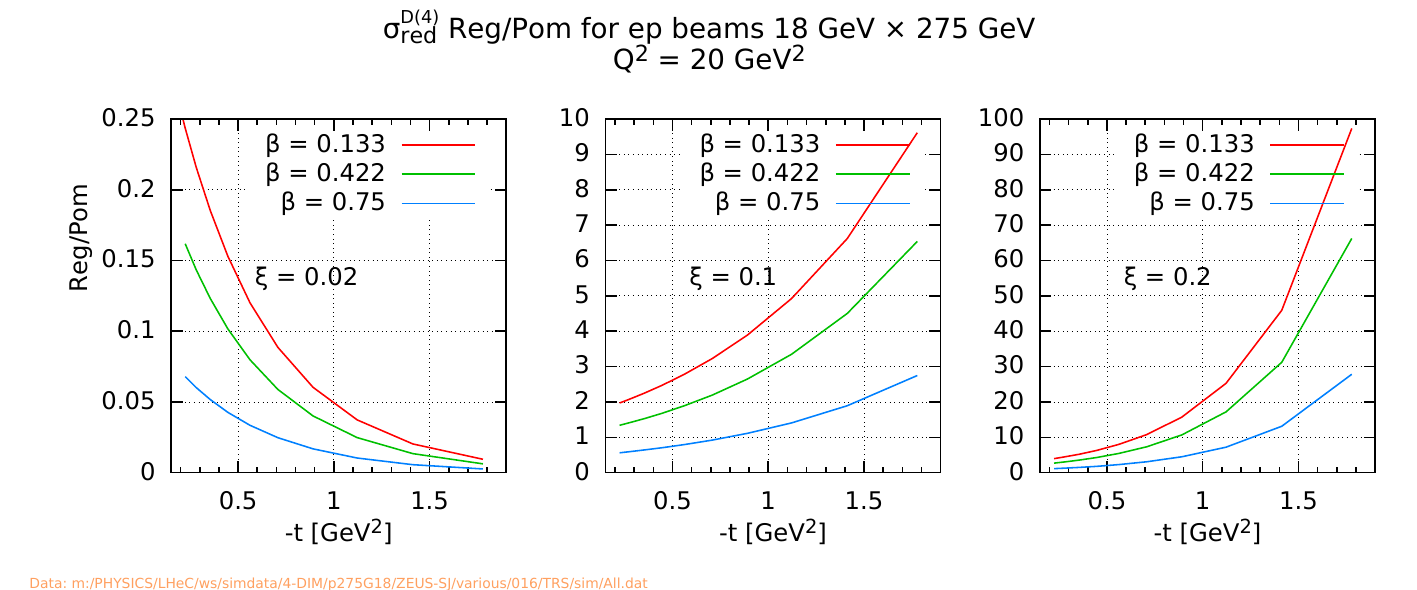}
}
\caption{Ratios of the Reggeon to Pomeron contributions to \srDD for the top energy setup and \(Q^2 = 20\,\GeV^2\), for selected values of $\beta,\xi$.}
\label{fig:ratioR2P}
\end{figure}

 As a final comment, we note that while we assume proton tagging at the EIC, the fit~\cite{Chekanov:2009aa} employed for generating the pseudodata considers only partially proton-tagged data -- although their compatibility with those extracted considering a rapidity gap  was established in the overlapping kinematic region. The DPDFs used for the pseudodata generation are corrected for the proton dissociation and thus normalized to the (non-dissociated) proton in the final state.

\subsection{Fit Methodology}
\label{subsec:fits}

In fitting the pseudodata, the 
Pomeron and Reggeon fluxes are parameterized according to ~\eqref{eq:fluxes}, with 3 parameters each, corresponding to $B$, $\alpha(0)$ and $\alpha^\prime$. %
 The parton densities for the Pomeron and Reggeon are 
 parameterized at an initial scale $\mu_0$ with a functional form 
\be
f^\mathbb{M}_k(x, \mu_0^2) = A^\mathbb{M}_k\, x^{B^\mathbb{M}_k}\, (1-x)^{C^\mathbb{M}_k}
(1 + D^\mathbb{M}_k\, x^{E^\mathbb{M}_k}),
\label{eq:initparm}
\ee
with $\mathbb{M}=\pom,\regg$. 
As in the fits performed by the HERA collaborations, we consider only $k=q,g$ with $q$ summed over all light quark and antiquark species\footnote{Quarks enter the inclusive cross section via their squared charge-weighted sum and, thus, different flavors cannot be resolved using inclusive neutral current
data alone.}. $D^\mathbb{M}_k$ is always set to 0 except for the Reggeon quark case, and $E^\mathbb{\regg}_q$ is fixed at $-1$,
such that there is a total of 13 parameters 
describing the Pomeron and Reggeon
PDFs, and a maximum of 19 free parameters to be fitted overall. These initial parametrizations are then evolved using the DGLAP evolution equations~\cite{Gribov:1972ri,Gribov:1972rt,Altarelli:1977zs,Dokshitzer:1977sg} at NLO. The initial scale $\mu_0^2$ is set to $1.8\,\GeV^2$ and, as in previous studies~\cite{Armesto:2019gxy,Armesto:2021fws}, no intrinsic heavy flavor is considered and  the charm and bottom quark DPDFs are generated radiatively via DGLAP evolution. 
The structure functions are calculated in a General-Mass Variable Flavor Number  scheme (GM-VFNS)~\cite{Collins:1986mp,Thorne:2008xf} which 
ensures a smooth transition of $F_\mathrm{2,L}$ across the flavor thresholds by including
$\mathcal{O}(m_h^2/Q^2)$ corrections. We fix $\alpha_s(M_Z) = 0.118$, $m_c =1.35$ GeV and $m_b = 4.3$ GeV for the fits, while the corresponding default values used in the GRV parametrizations are used for the generation of the pseudodata.

Our fitting model is different from that used to generate the pseudodata in respect to the Reggeon parametrization
where the GRV-like Reggeon includes
several separate quark species, inherited from the pion.
We have checked for bias due to this difference, by comparing
results of fits to
the pseudodata at the level of the $t$-integrated PDFs
\be
f_k^{D(3)}(z, Q^2, \xi) =
\Phi_\pom(\xi)\, f^\pom_k(z, Q^2)
+
\Phi_\regg(\xi)\, f^\regg_k(z, Q^2) \ .
\label{eq:D3pdfs}
\ee
The results, including decomposition into 
Pomeron and Reggeon contributions,
are shown in Figs.~\ref{fig:pdfs_xi0.01} and~\ref{fig:spdfs_xi0.1}.
Our model \Eq{eq:initparm}, labeled `Parameterized` (continuous lines),
is compared to the `GRV Reggeon` \Eq{eq:grvpdfs} (dashed lines). The latter depicts merely the input pseudodata.
As expected, due to the same
treatment in the pseudodata and the fits, the agreement for the
Pomeron is excellent.
There are some minor differences for 
both the quark and the gluon contributions in the
the Reggeon case, particularly 
at larger $\xi$, though the overall input shape
is reproduced reasonably well.

\begin{figure}[htb!]
\centerline{
\includegraphics[width=0.8\columnwidth,clip,trim=0 20 0 33]{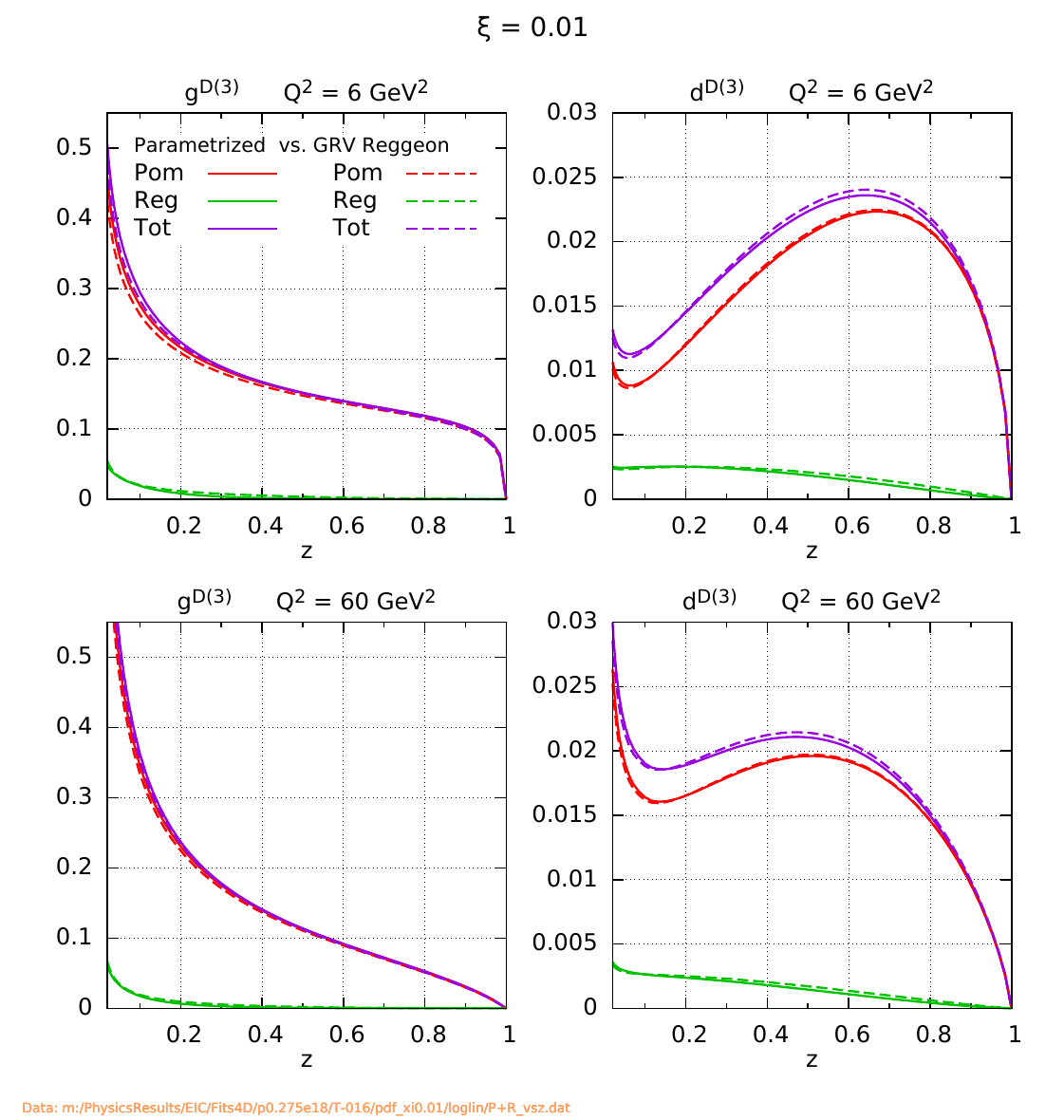}
}
\caption{$t$-integrated DPDFs~\eqref{eq:D3pdfs} versus 
momentum fraction $z$ for gluons (left plots) and quarks (right plots) at $Q^2=6$ (upper plots) and 60 (lower plots) GeV$^2$, for $\xi=0.01$. 
The solid lines are the results of the fits with our 
parametrization~\eqref{eq:initparm}, while the dashed lines correspond to the input pseudodata, using the GRV set for the Reggeon. Contributions from Reggeon, Pomeron and their sum are shown in green, red and magenta respectively.}
\label{fig:pdfs_xi0.01}
\end{figure}

\begin{figure}[htb!]
\centerline{
\includegraphics[width=0.8\columnwidth,clip,trim=0 20 0 33]{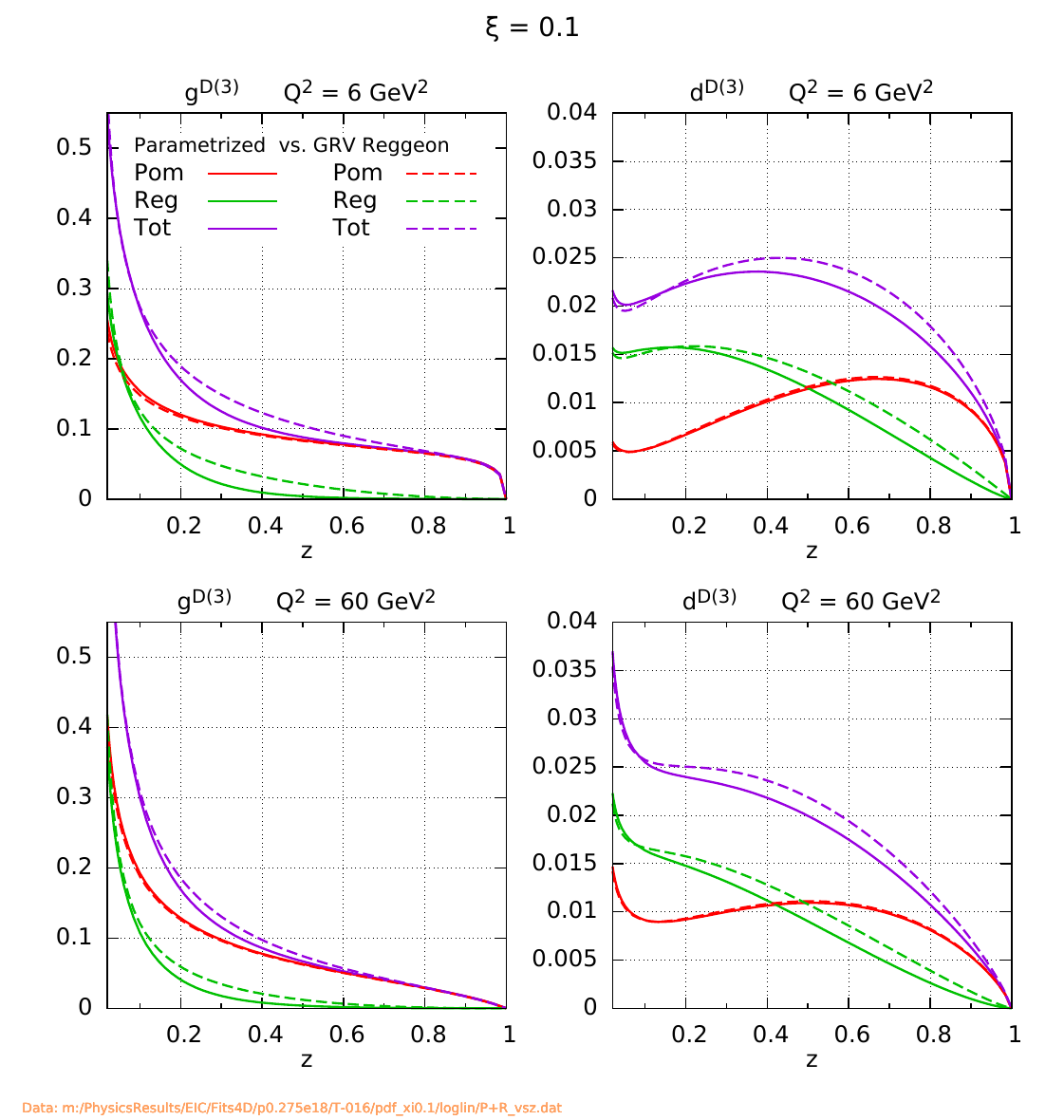}
}
\caption{As for Fig.~\ref{fig:pdfs_xi0.01} but for $\xi=0.1$.}
\label{fig:spdfs_xi0.1}
\end{figure}

Our fitting strategy at the top energy is as follows. 
Since the Pomeron contribution dominates at large $-t$ and small $\xi$,
see~\Fig{fig:ratioR2P},
a Pomeron-only fit is first performed in the
region $-t  \in  [0.4, 2] \GeV^2$ and $\xi \in  [0.0004, 0.007]$. The parameters extracted from this fit are then taken as initial values for 
a complete Pomeron + Reggeon fit over the full accessible kinematic range.
At the lower energy, where the Reggeon contribution 
dominates in the full accessible kinematic region,
our strategy is to rely on the precise knowledge 
of the 
Pomeron fluxes and PDFs from HERA, fixing them to the results 
from~\cite{Chekanov:2009aa} and fitting only the Reggeon parameters.

For the uncertainties in the extracted parameters, the standard Hessian method~\cite{Pumplin:2001ct} is used. These uncertainties, corresponding to $1\sigma$ (68\% c.l.), are then propagated into the Pomeron and Reggeon PDFs. As commented previously, 
a fully correlated normalization uncertainty 
of 2\% is also considered.
As an additional check, we also compare the fit results for different Monte Carlo replicas of the pseudodata with independently performed smearing.

\section{Results}
\label{sec:results}

We first consider the high energy scenario $E_e=18 \, {\rm GeV} \times E_p=275 \, {\rm GeV}$ with high luminosity ${\cal L}=100 \, {\rm fb}^{-1}$.

In Figs.~\ref{fig:cmpEB_Pomeron_xi} and \ref{fig:cmpEB_Reggeon_xi} we show the relative uncertainties on the gluon and quark components of the leading (Pomeron) and secondary (Reggeon) exchanges.
The uncertainties are extracted from fits including 
all data up to $-t_{\rm min}=1.5 \, \rm GeV^2$. 
The statistical errors above that value are rather large due to the steeply falling cross section.
The uncertainties are presented as a function of $z$ for fixed values of $\xi=0.01$ and $0.1$ at $Q^2=6 \, \rm GeV^2$ and
$60 \, \rm GeV^2$. 
The error sources that are assumed not to be correlated between data
points lead to uncertainties on the 
DPDFs that vary with the kinematics. On the other 
hand, for a single beam energy, the 
data error sources that are assumed to be fully correlated 
between data points 
propagate directly to normalization uncertainties in the DPDFs. To
illustrate their potential impact, a normalization uncertainty of 2\%
is added in quadrature with the error bands in the figures,
leading to the full uncertainty ranges 
shown as red dashed lines.

In general, the uncertainties on the Pomeron gluon and quark are very small,
often 
below the $1 \%$ level before taking the normalization uncertainty into
account. The normalization uncertainty is dominant throughout the phase
space, though it is likely to be possible to reduce it 
with the inclusion of
data from different beam energy
configurations. 
The uncertainty bands arising from uncorrelated sources grow at small $z$,
where the EIC acceptance at low $x$ is the limiting factor.
To illustrate better the behavior at low $z$ we show the uncertainties on the Pomeron and Reggeon contribution using a logarithmic $z$ scale in Figs.~\ref{fig:scmpEB_Pomeron_log_xi} and \ref{fig:scmpEB_Reggeon_log_xi} respectively.
The uncertainties grow slowly as $z$ decreases, down to the $10^{-3}$ which is the approximate kinematic limit of the data.
The Pomeron gluon density uncertainty also increases
at large $z$, particularly for the gluon density, which lacks direct
constraints from inclusive data. The high $z$ gluon precision
might be improved by including diffractive jet data, as was the case at HERA.

The Reggeon contribution has uncertainties 
below the $2 \%$ level for the quark case except for very large $z$. 
On the other hand, the gluon component of the Reggeon has larger uncertainties, particularly at moderate $z$. 
However, as is evident from Figs.~\ref{fig:pdfs_xi0.01} 
and~\ref{fig:spdfs_xi0.1} (left panels) the Regggeon gluon 
density is itself 
rather small away from the smallest $z$ values, so the absolute size
of the uncertainty is not large. 
In the Reggeon case, the 
2\% normalisation uncertainty is dominant for the quark density, but its 
influence is relatively small for the gluon case.

In Figs.~\ref{fig:cmpEB_Pomeron_xi} and \ref{fig:cmpEB_Reggeon_xi} the sensitivity to the maximum value of $\xi$ in the measurement 
is also assessed by imposing various  cuts, 
$\xi < \xi_{\rm max} = 0.15,0.22,0.3$. The corresponding numbers of simulated data points are $N=7807, 8367$ and 
$8835$. We observe some dependence on $\xi_{\rm max}$ for the uncertainties in the case of the Reggeon gluon contribution, and somewhat smaller sensitivity for the Reggeon quark component. In the case of the Pomeron exchange the uncertainties are practically insensitive to the variation of this cut, as expected since the Pomeron is dominant 
only at low values of $\xi$.

In Figs.~\ref{fig:cmpEB_Pomeron_t} and~\ref{fig:cmpEB_Reggeon_t}
the uncertainty bands are shown for different 
assumed measurement ranges in $t$ such that
the minimum value is $t_{\rm min} = -0.5 \, \rm GeV^2$, 
$-1.0 \, \rm GeV^2$ and
$-1.5 \, \rm GeV^2$, shown as red, green and blue regions respectively.
In each case, $\xi_{\rm max}$ is set to $0.22$.
The Pomeron contribution is not very sensitive to this 
variation. The Reggeon uncertainty exhibits some dependence, 
but it is less important than the $\xi_{\rm max}$ choice. 

Altogether, these results 
show that an excellent precision is achievable simultaneously for  
the Reggeon and Pomeron contributions,
even with a relatively unaggressive kinematic range 
choice such as $t > -0.5 \, \rm GeV^2$ and 
$\xi < 0.15$.

Although not shown, we have also studied the influence on the results 
when changing from the dense binning scheme used here 
to a sparse binning scheme. 
The differences are very small except for the 
case of the Reggeon gluon, whose small absolute 
values require a dense binning at large $\xi$ for a 
precise extraction.
We have also investigated the influence of
lowering the luminosity to 10 \ifb, which is 
negligible since the measurements are expected to be
strongly systematics dominated at the EIC.

\begin{figure}[htb]
\centerline{
\includegraphics[width=0.49\columnwidth,clip,trim=0 30 0 30]{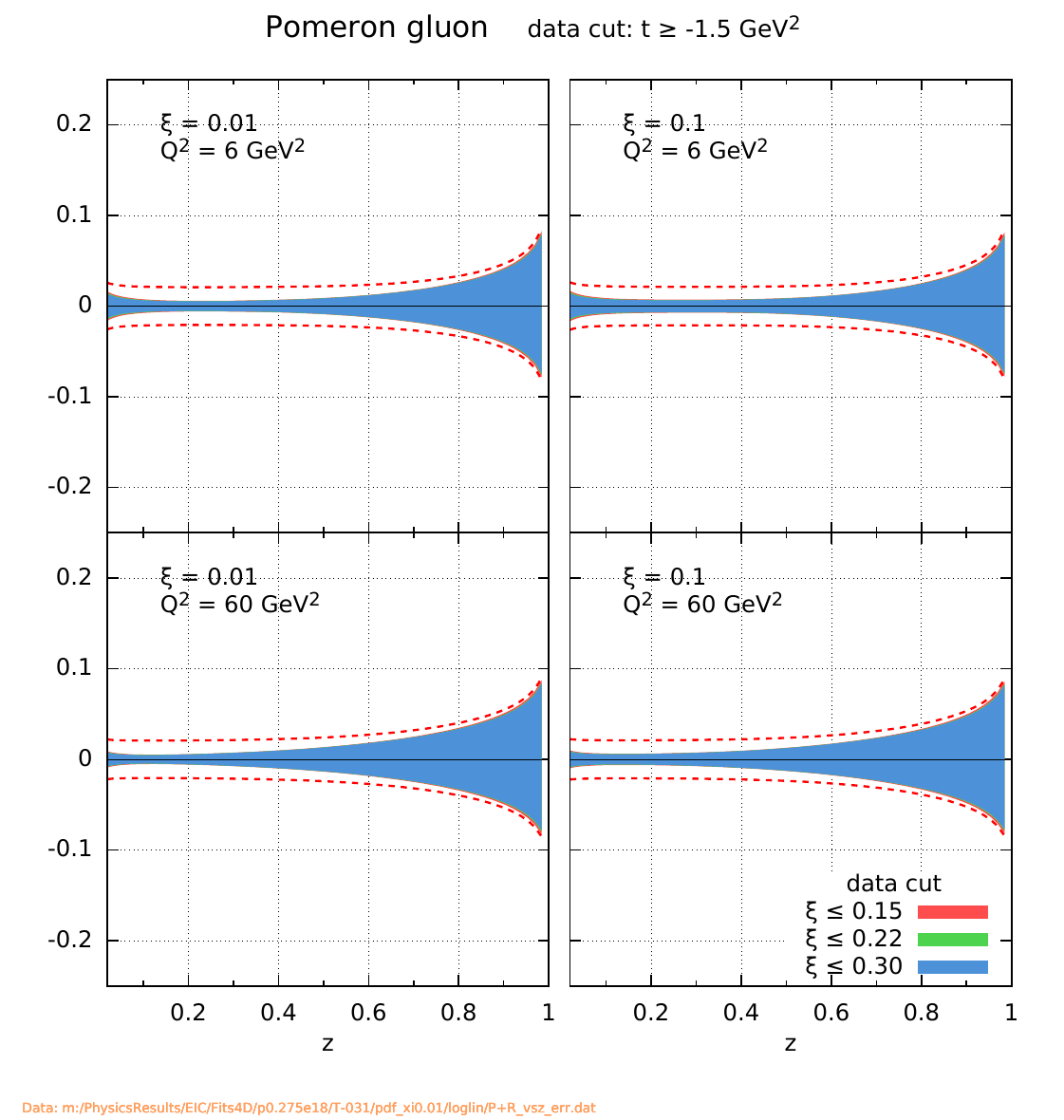}
\includegraphics[width=0.49\columnwidth,clip,trim=0 30 0 30]{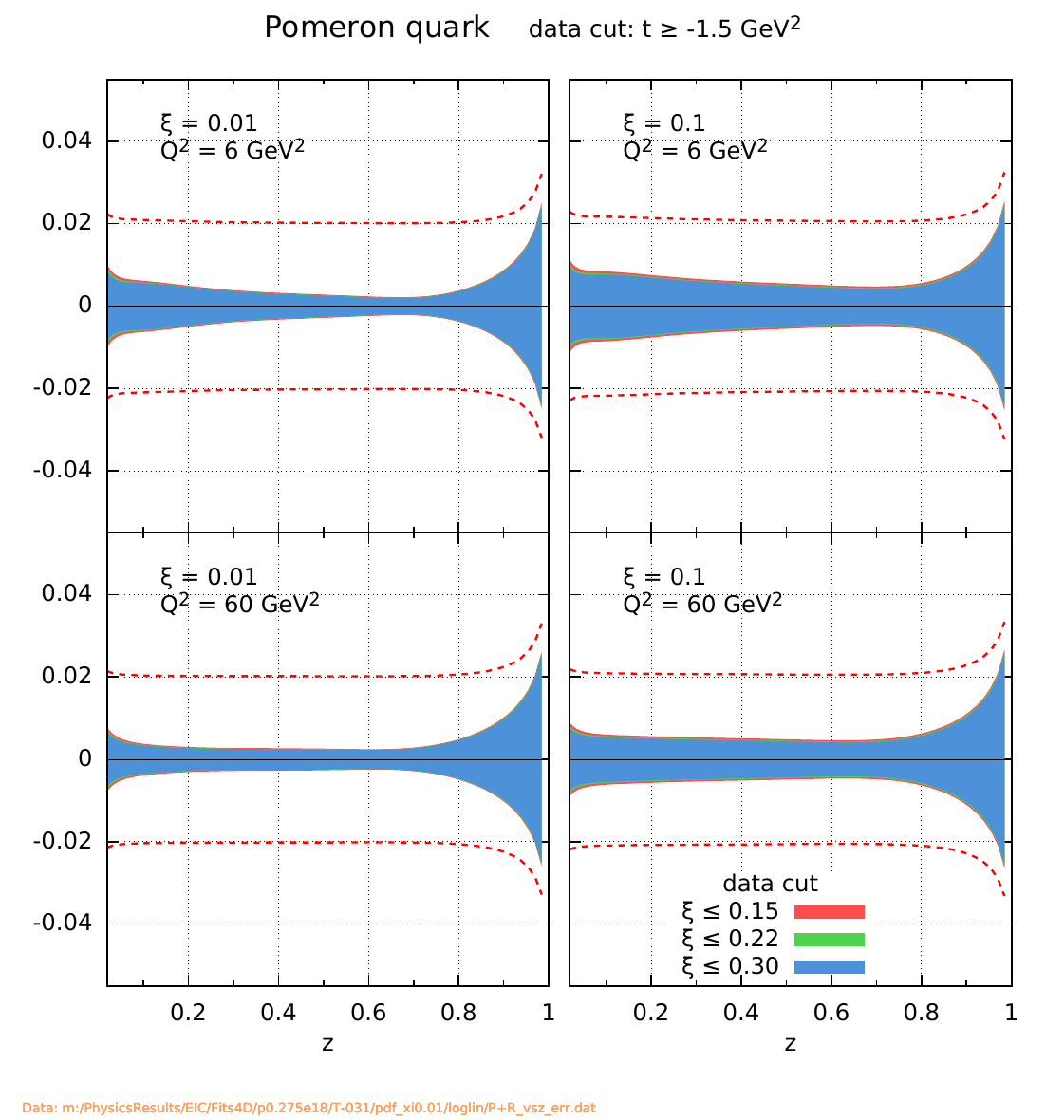}
}
\caption{Relative uncertainties for the  gluon (left) and quark (right) contributions to the Pomeron
PDFs as a function of longitudinal momentum fraction $z$ for fixed values of $\xi=0.01,0.1$ and $Q^2 =6,60 \,\GeV^2$.  Results 
are shown for three different cuts on $\xi<0.15,0.22,0.30$, corresponding to red, green and blue regions respectively. The range in momentum transfer is $-t \le 1.5 \,\GeV^2$.
The dashed lines depict error band limits upon including a normalization error of 2\% for the case of the lowest $\xi$ cut.}
\label{fig:cmpEB_Pomeron_xi}
\end{figure}

\begin{figure}[htb]
\centerline{
\includegraphics[width=0.49\columnwidth,clip,trim=0 30 0 30]{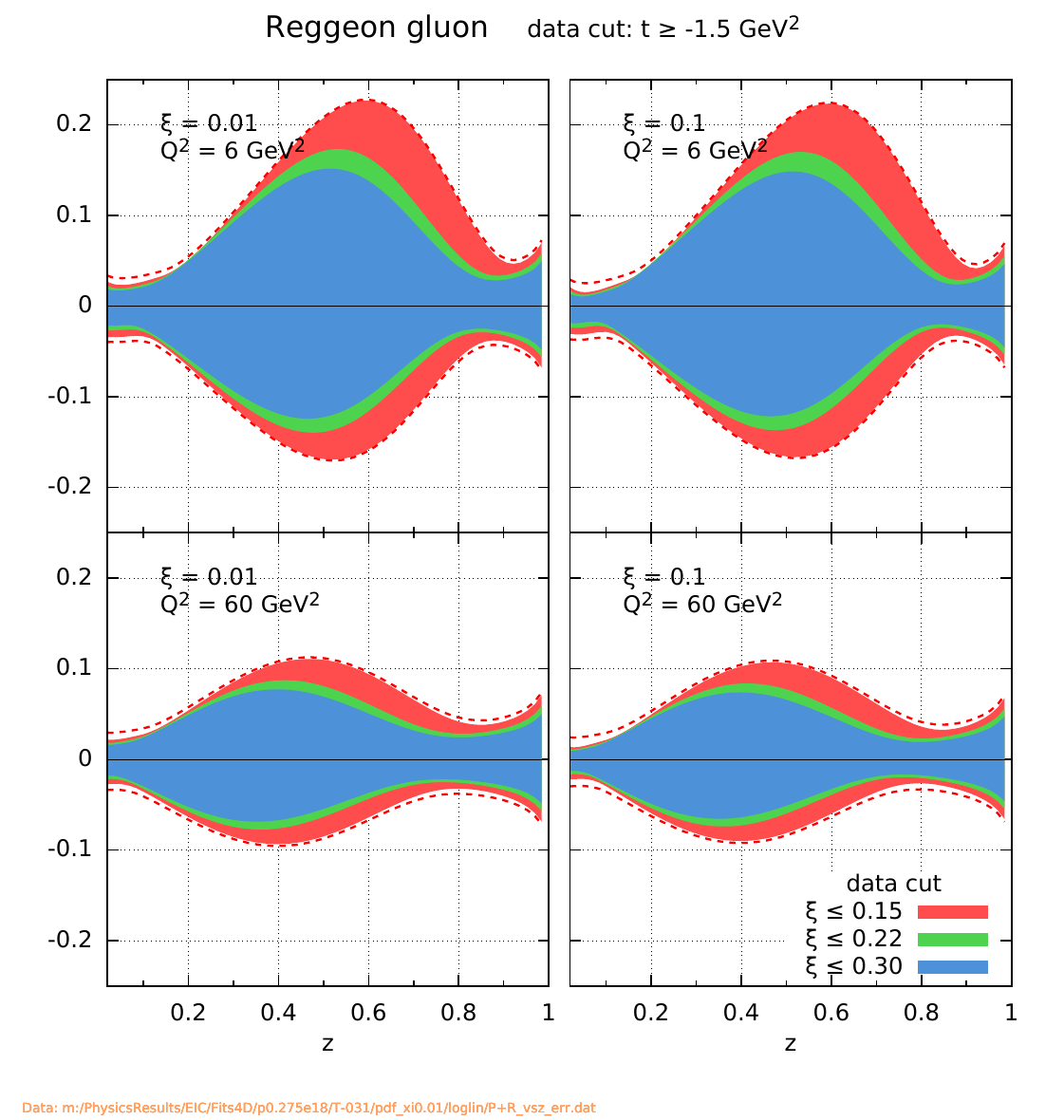}
\includegraphics[width=0.49\columnwidth,clip,trim=0 30 0 30]{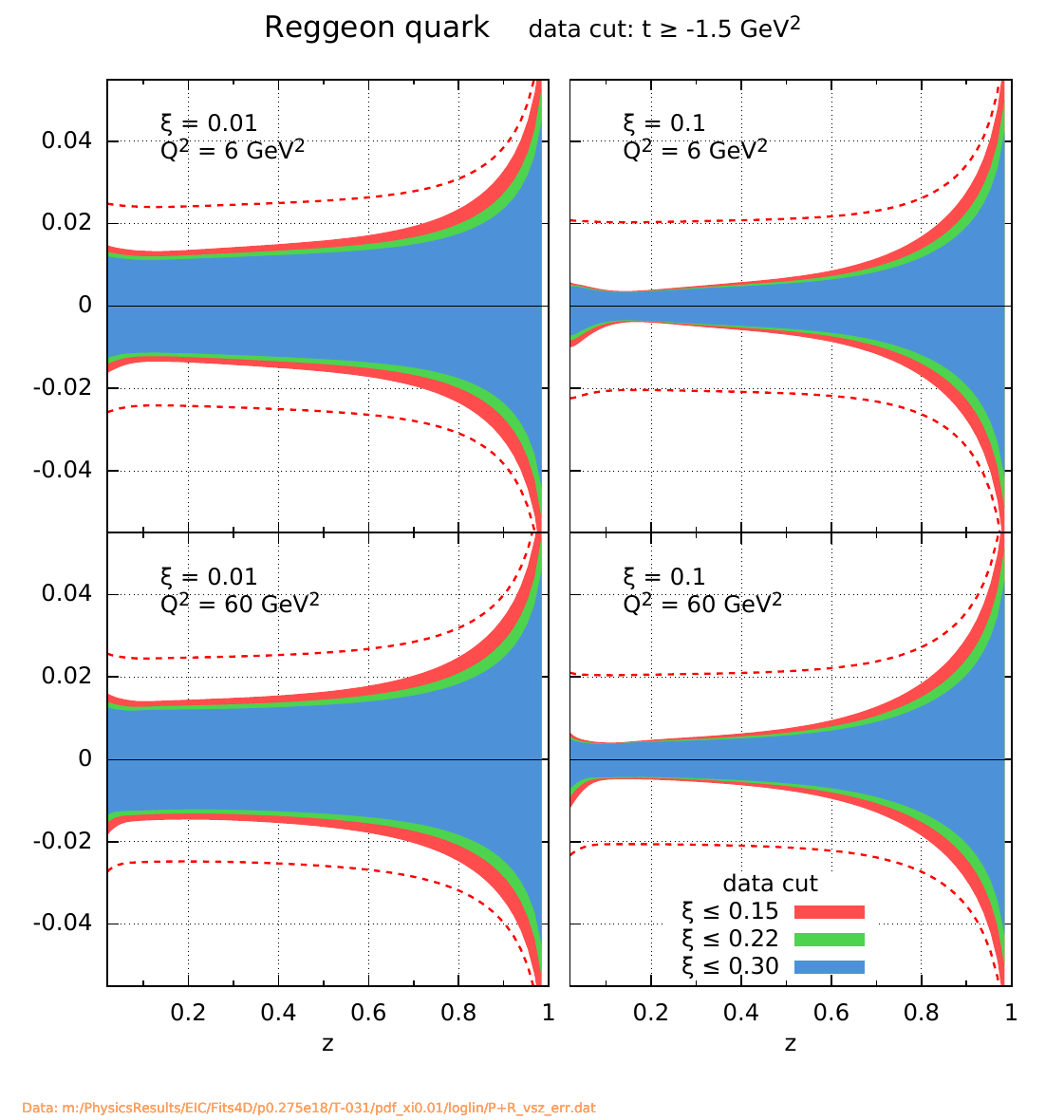}
}
\caption{Relative uncertainties for the  gluon (left) and quark (right) contributions to the Reggeon
PDFs as a function of longitudinal momentum fraction $z$ for fixed values of $\xi$  and $Q^2$.  The numerical values for the $\xi$ and $Q^2$ bins and $\xi$ cuts are, as well as the conventions for regions and lines, the same as  for Fig.~\ref{fig:cmpEB_Pomeron_xi}.}
\label{fig:cmpEB_Reggeon_xi}
\end{figure}

\begin{figure}[htb]
\centerline{
\includegraphics[width=0.49\columnwidth,clip,trim=0 30 0 30]{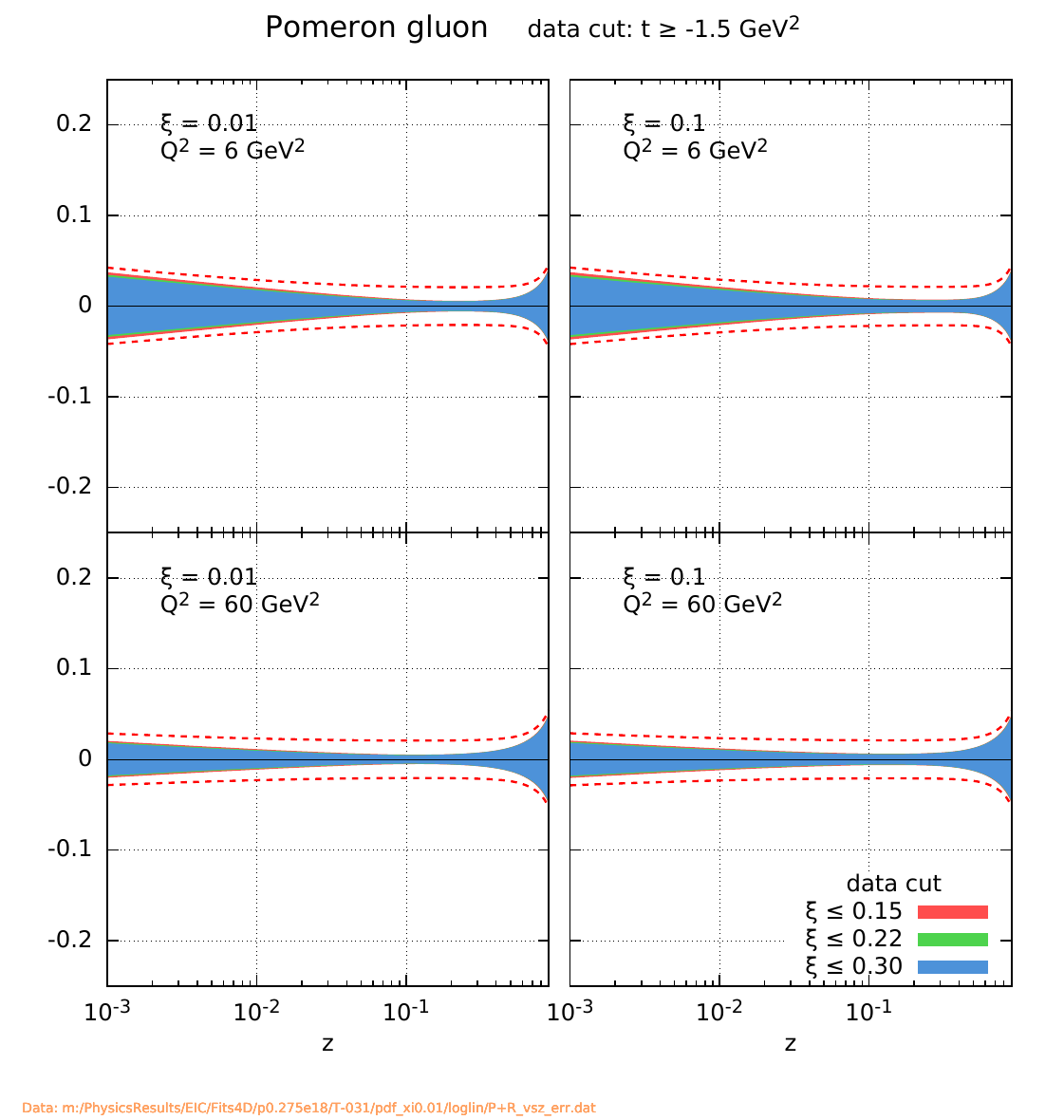}
\includegraphics[width=0.49\columnwidth,clip,trim=0 30 0 30]{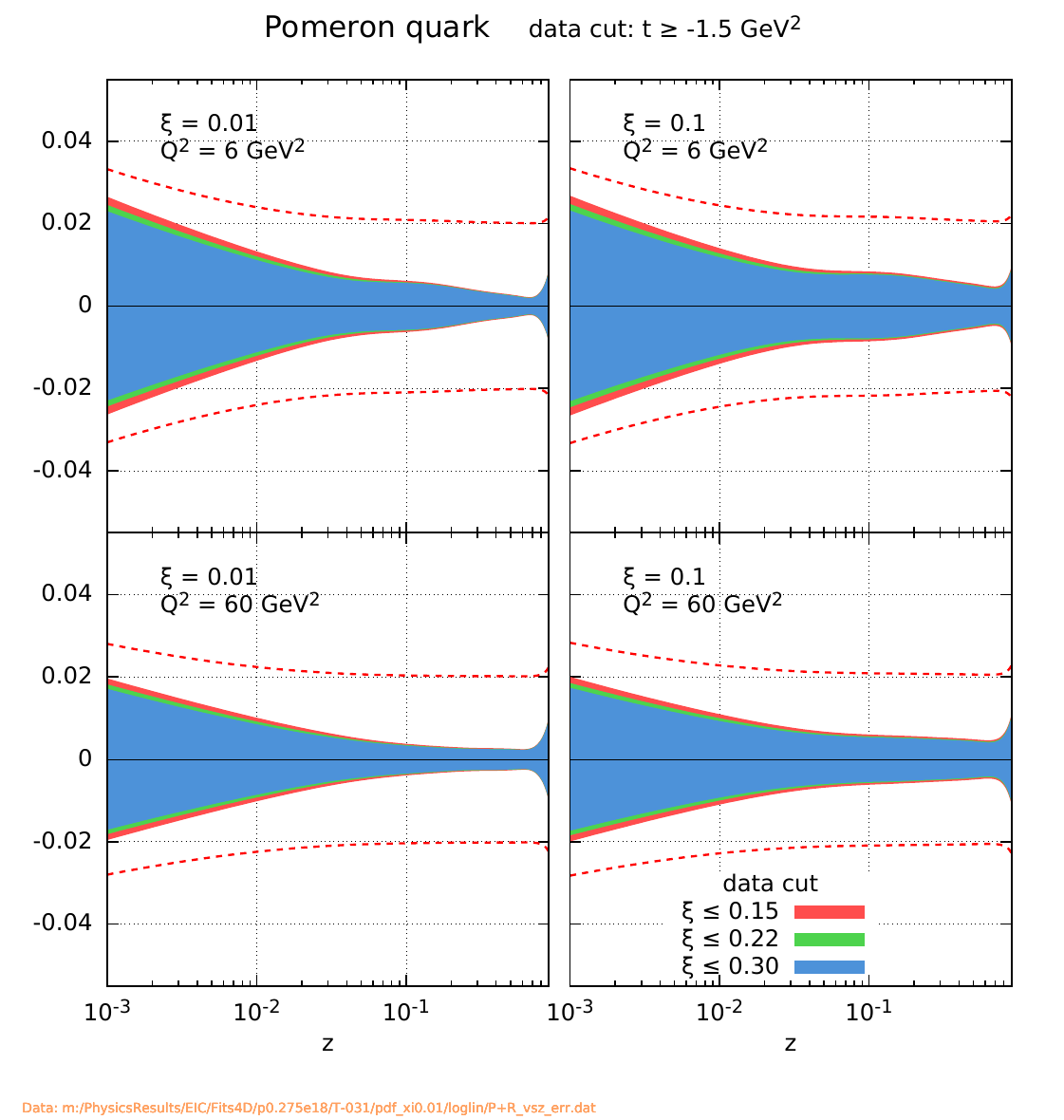}
}
\caption{The same as Fig.~\ref{fig:cmpEB_Pomeron_xi} but with  a logarithmic $z$ scale.}
\label{fig:scmpEB_Pomeron_log_xi}
\end{figure}

\begin{figure}[htb]
\centerline{
\includegraphics[width=0.49\columnwidth,clip,trim=0 30 0 30]{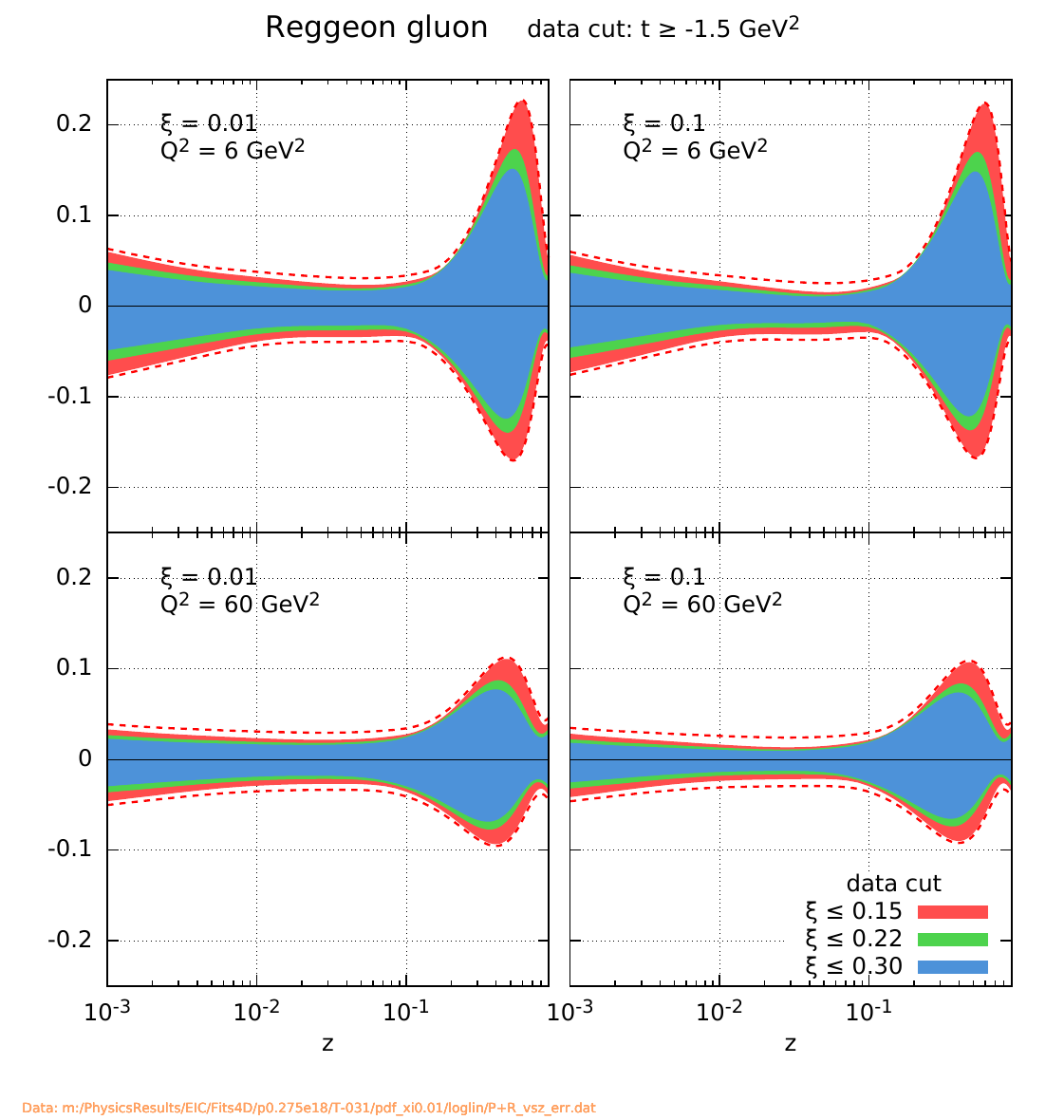}
\includegraphics[width=0.49\columnwidth,clip,trim=0 30 0 30]{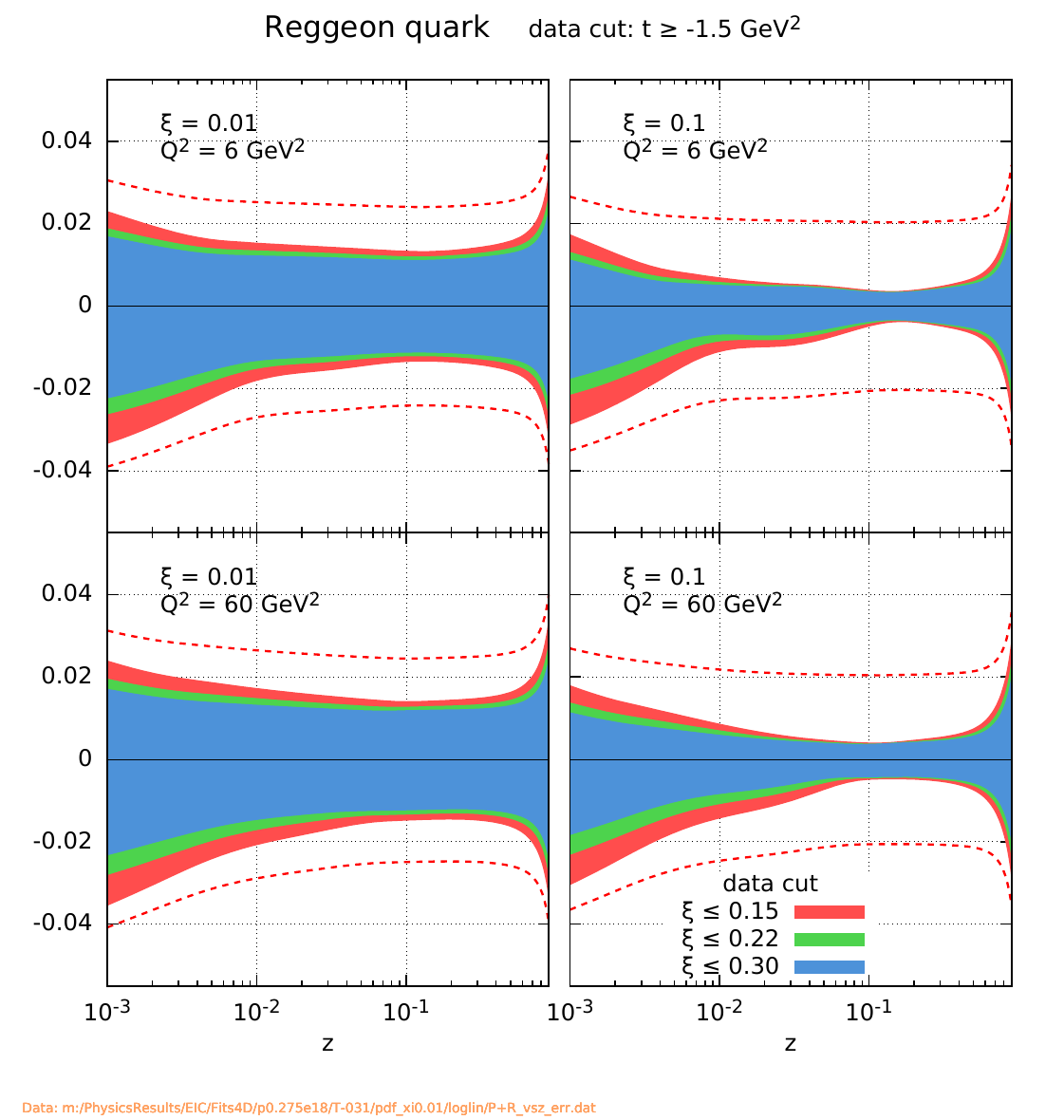}
}
\caption{The same as Fig.~\ref{fig:cmpEB_Reggeon_xi} but with a logarithmic $z$ scale.}
\label{fig:scmpEB_Reggeon_log_xi}
\end{figure}

\begin{figure}[htb]
\centerline{
\includegraphics[width=0.49\columnwidth,clip,trim=0 30 0 30]{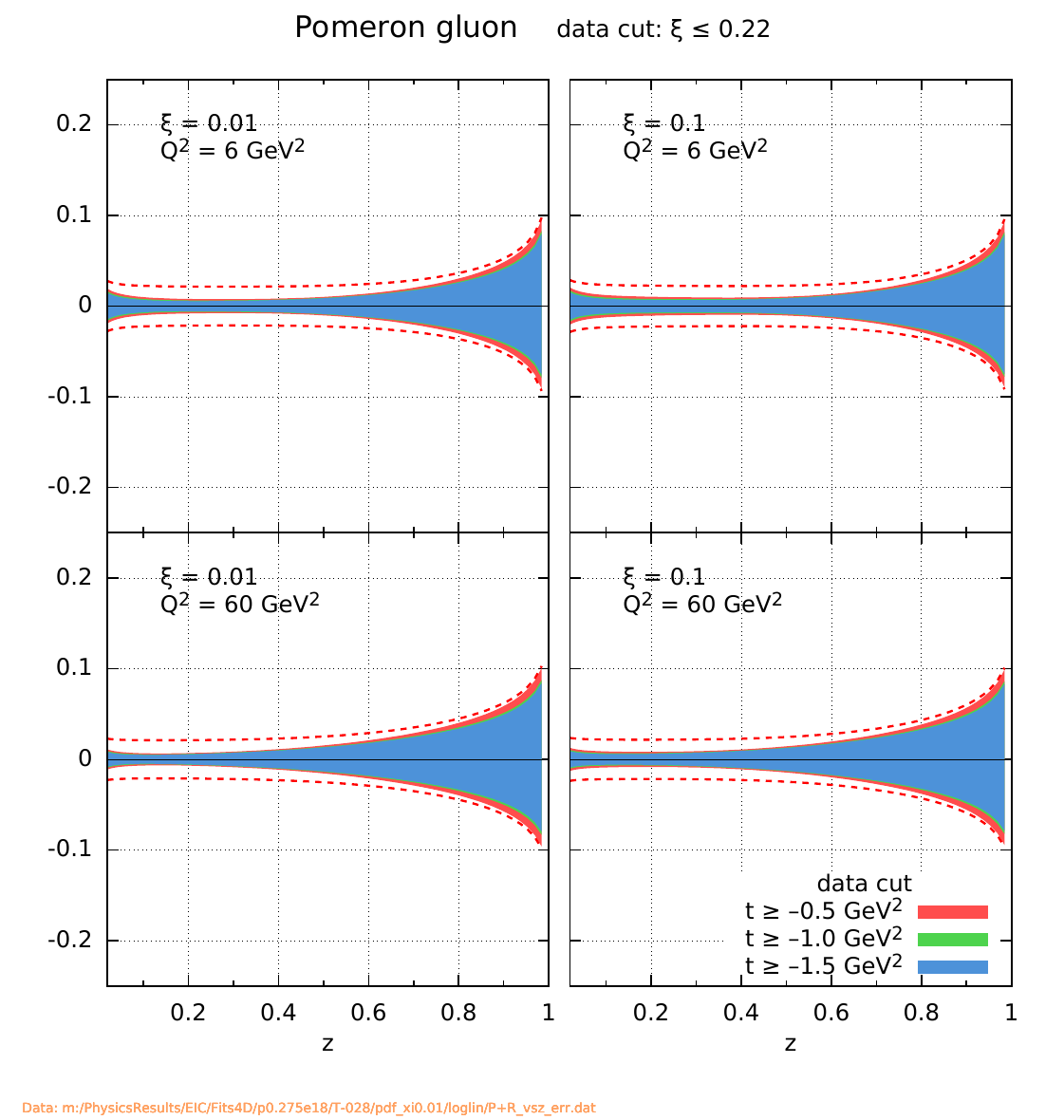}
\includegraphics[width=0.49\columnwidth,clip,trim=0 30 0 30]{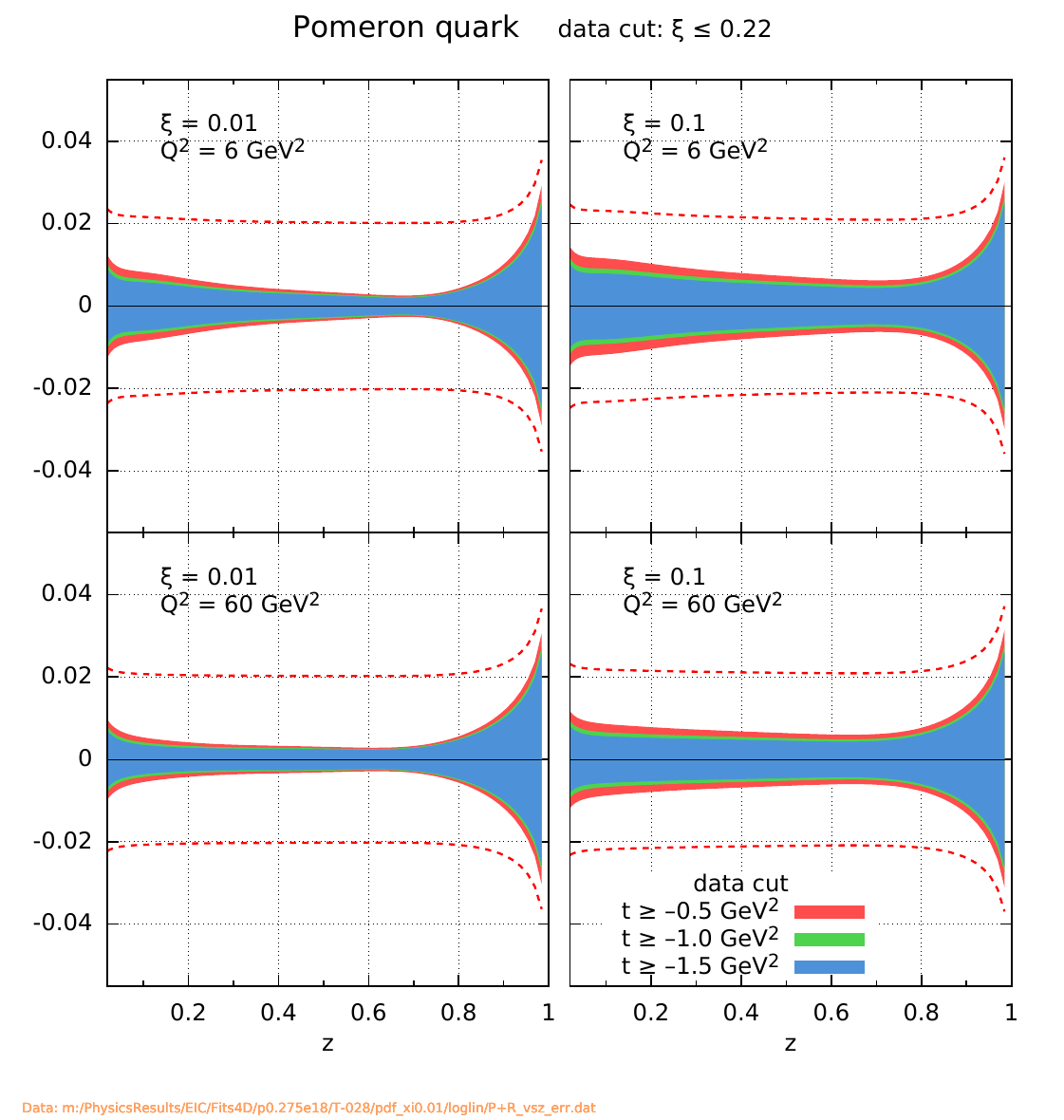}
}
\caption{Relative uncertainties for the  gluon (left) and quark (right) contributions to the Pomeron PDFs, as a function of longitudinal momentum fraction $z$ for fixed values of $\xi=0.01,0.1$ and $Q^2 =6,60 \, \rm GeV^2$.
Results are shown for data cuts: $t \ge -0.5, -1.0, -1.5 \,\GeV^2$, corresponding to the red, green and blue regions, respectively, with
$\xi_{\rm max} = 0.22$ in all cases.  
The dashed lines depict error band limits upon including a normalization error of 2\% for the case of the lowest $-t$ cut.}
\label{fig:cmpEB_Pomeron_t}
\end{figure}

\begin{figure}[htb]
\centerline{
\includegraphics[width=0.49\columnwidth,clip,trim=0 30 0 30]{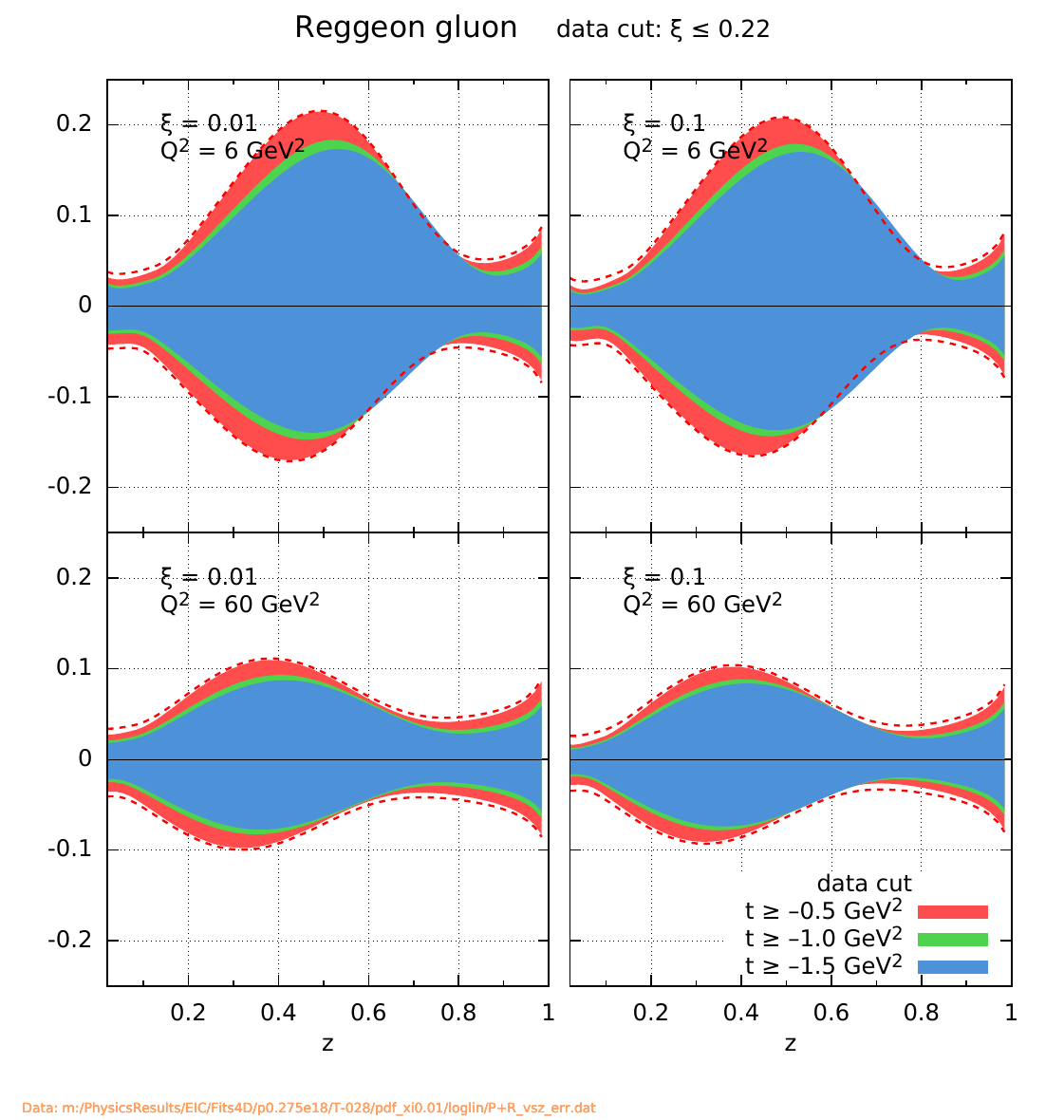}
\includegraphics[width=0.49\columnwidth,clip,trim=0 30 0 30]{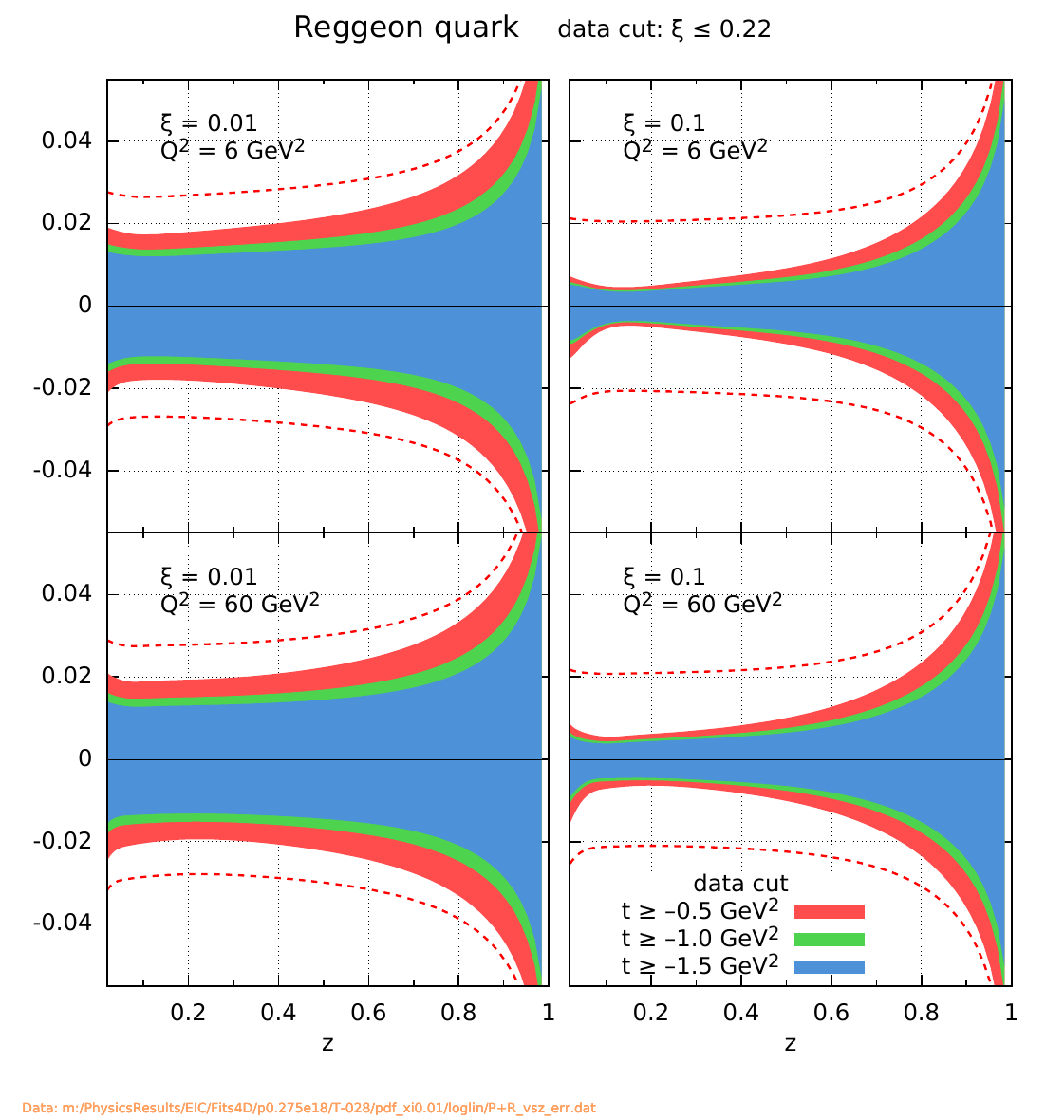}
}
\caption{As for Fig.~\ref{fig:cmpEB_Pomeron_t} but for the Reggeon.}
\label{fig:cmpEB_Reggeon_t}
\end{figure}

We have also studied a lower EIC beam energy scenario, 
with $E_e=5 \, {\rm GeV} \times E_p=41 \, {\rm GeV}$ 
and a limited luminosity of ${\cal L}=10$ \ifb.
In this case, the minimum kinematically accessible 
$\xi$ value
is $\sim  0.01$, meaning that the sensitivity to the Pomeron 
contribution is severely restricted. As discussed in 
section~\ref{subsec:fits}, we therefore adopt a 
different fitting strategy for this scenario
in which the Pomeron contribution is fixed 
by the results from HERA data, and 
the fit is performed only to the Reggeon contribution. 

In Fig.~\ref{fig:spdfs_lowE_Reggeon} we show the extracted DPDFs for the 
quark and gluon contributions to the
Reggeon structure at two values of $\xi=0.01,0.1$ and 
two values of $Q^2=6, 60$ GeV$^2$, while in Fig.~\ref{fig:spdfsR_lowE_Reggeon} we show the 
corresponding relative uncertainties.
The results are not very sensitive to 
variations of the $-t_{\rm min}$ 
restriction between $1.8$ and 
$1.2$ GeV$^2$, and acceptable fits can 
even be obtained with $-t_{\rm min}=0.6$ GeV$^2$. On the 
other hand, the range in $\xi$ is very important,
as illustrated by the colored bands, 
which show the uncertainty variations when 
choosing  $\xi_{\rm max}$ to be $0.33,0.22$ or $0.15$,
with $-t_{\rm min}=1.2 $ GeV$^2$ in each case. 
The numbers of data points for these 
scenarios are $948, 816 $ and $ 672$, respectively. 
 The results of the fits indicate that a good precision 
 on the quark contribution to the Reggeon is 
 achievable for the low beam energy scenario,
 particularly if the $\xi$ range 
 extends to at least $0.22$. However, a wider $\xi$ range
 is required to obtain good constraints on the gluon contribution, 
 ideally extending as far as $0.33$.

\begin{figure}[htb]
\centerline{
\includegraphics[width=0.49\columnwidth,clip,trim=0 20 0 33]{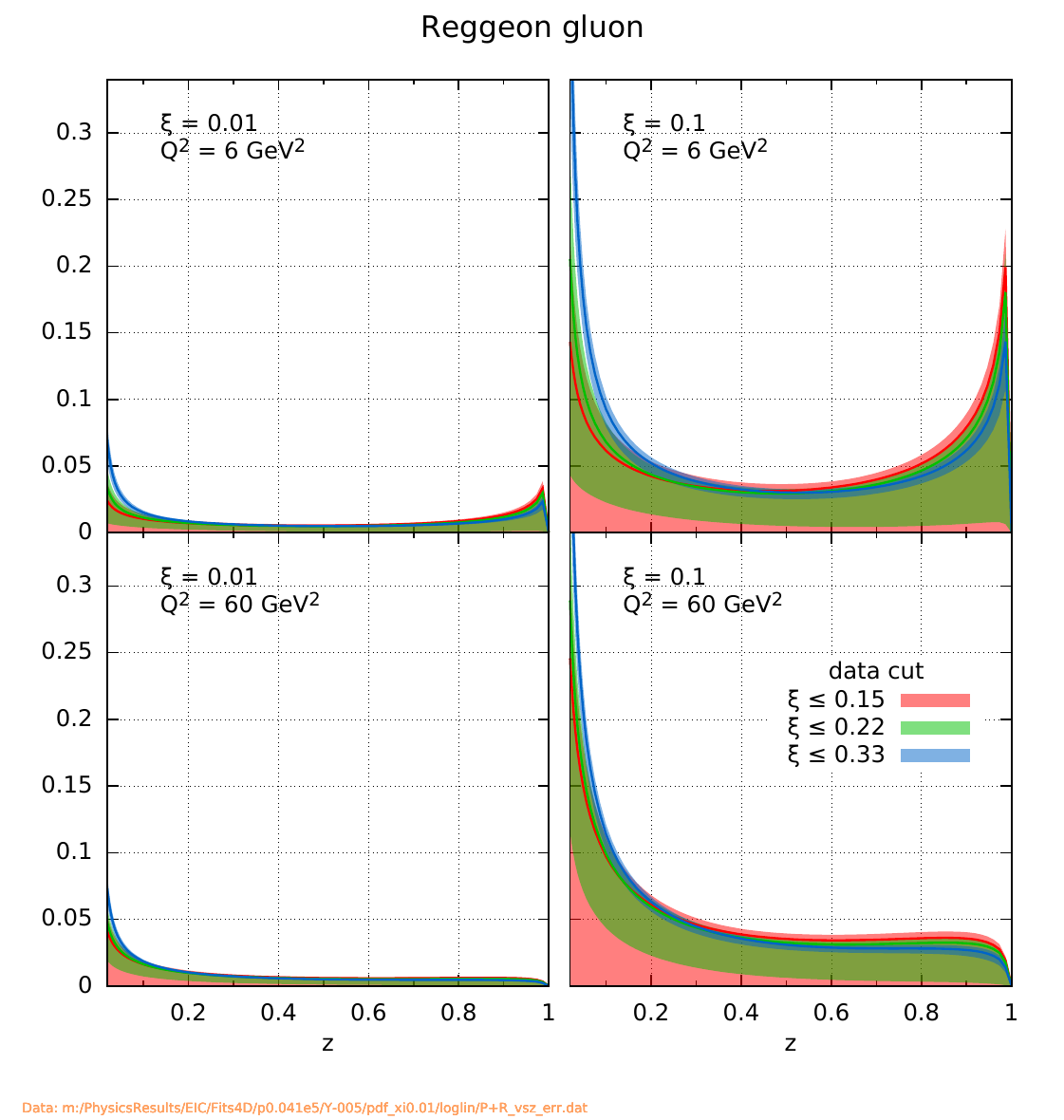}
\includegraphics[width=0.49\columnwidth,clip,trim=0 20 0 33]{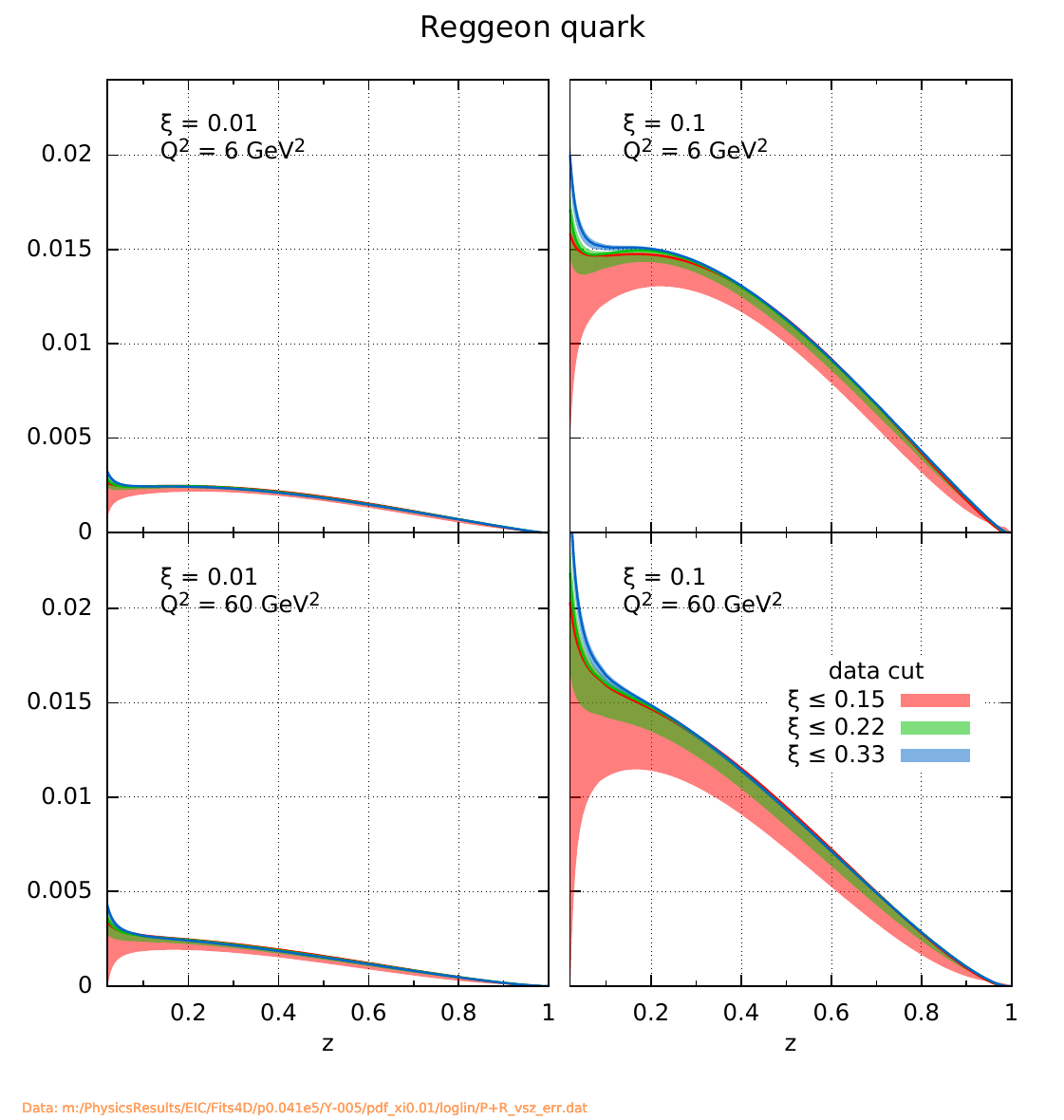}}
\caption{$t$-integrated DPDFs~\eqref{eq:D3pdfs} for the Reggeon versus $z$ for gluons (left plots) and quarks (right plots) at $Q^2=6$ (upper plots) and 60 (lower plots) GeV$^2$, in the lower beam energy configuration,
$E_e \times E_p = 5 \times 41$ GeV. The two columns in each panel correspond to   $\xi=0.01$ (left) and $\xi=0.1$ (right). 
Uncertainties are shown for three different  scenarios 
with $\xi_{\rm max} = 0.15$ (orange), $0.22$ (green) or $0.33$ (blue),
with $t_{\rm min} = -1.2 \ {\rm GeV^2}$ in each case. }
\label{fig:spdfs_lowE_Reggeon}
\end{figure}

\begin{figure}[htb]
\centerline{
\includegraphics[width=0.49\columnwidth,clip,trim=0 20 0 33]{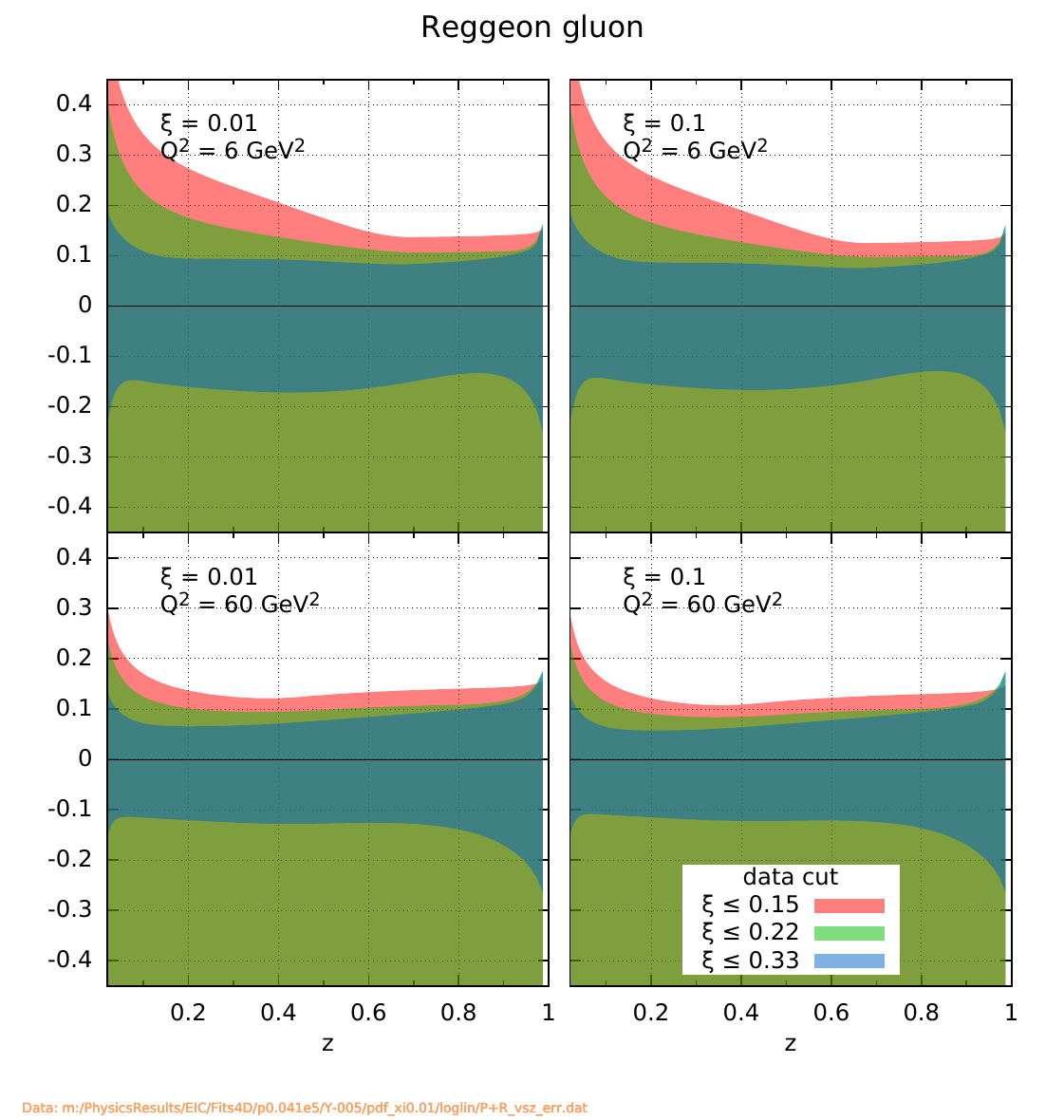}
\includegraphics[width=0.49\columnwidth,clip,trim=0 20 0 33]{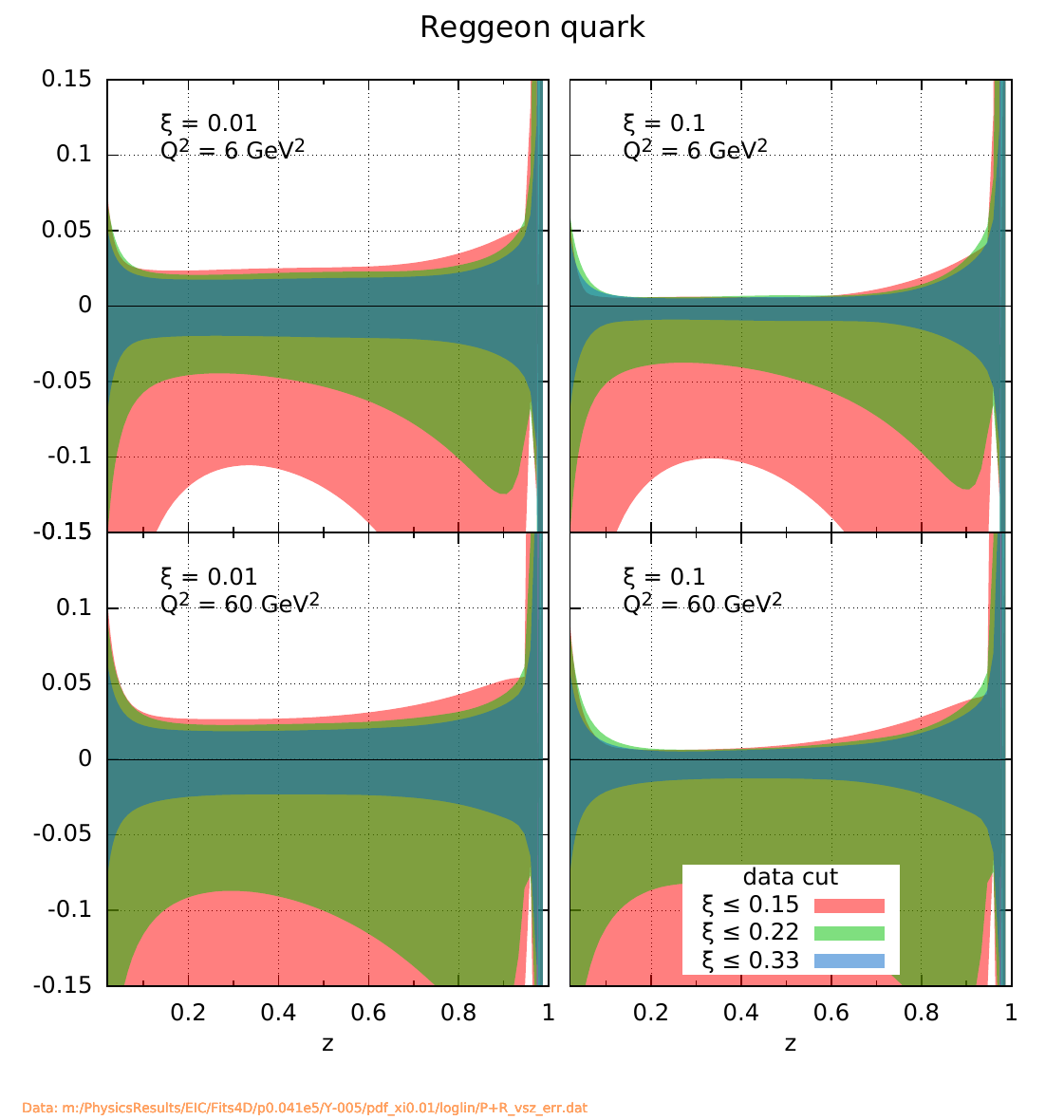}}
\caption{Relative uncertainties for the  gluon (left) and quark (right) contributions to the Reggeon
PDFs as a function of longitudinal momentum fraction $z$ for fixed values of $\xi=0.01,0.1$ and $Q^2 =6,60 \,\GeV^2$, in the lower beam energy configuration,
$E_e \times E_p = 5 \times 41$ GeV.  Results 
are shown for three different cuts on $\xi<0.15,0.22,0.30$, corresponding to red, green and blue regions respectively. The range in momentum transfer is $-t \le 1.2 \,\GeV^2$.}
\label{fig:spdfsR_lowE_Reggeon}
\end{figure}

For both beam energy configurations, 
we have checked 
the impact of choosing different Monte Carlo 
replicas for the pseudodata, by performing separate fits for
independently generated samples. The differences in fit quality and in the extracted uncertainties are marginal and do not affect the conclusions. 

As a final comment, in real data, final states with low 
$M_X$ may be produced by other mechanisms
than DIS from a composite Pomeron and Reggeon with
universal partonic structure,
for example, models with perturbative exchanges at higher twist
that are generally successful in describing exclusive vector meson production.
In the pseudodata simulated for this study,
we have not implemented any other mechanisms, so going down to $M_X \approx 1.2\,\GeV$ does not spoil the fitting procedure.
However, 
a cut of, e.g., $M_X > 2\,\GeV$ 
is commonly included in DPDF fits, so
we have checked how selecting only data with $M_X > 2\,\GeV$ influences the 
results.
The number of data points goes down by ~10\% for the high energy scenario, and by ~20\% for the low one.
The resulting DPDFs, \chiNDF values and uncertainties are very similar
to those shown here.


\section{Summary and Outlook}
\label{sec:conclu}

In this work we have analyzed the possibility for the simultaneous extraction of the partonic content of leading (Pomeron) and sub-leading (Reggeon) colorless exchanges in diffractive DIS off protons at the EIC. For this purpose, we have generated pseudodata in the four diffractive kinematic variables $\beta,Q^2,\xi,t$ using a model which describes available HERA data. We have considered different binnings and Monte Carlo samples, focusing on cases with one year's integrated luminosities at the highest or lowest 
center-of-mass energies. We have made conservative assumptions on the systematic uncertainties, and studied the influence of different cuts on the kinematic variables.
With these pseudodata, we have made fits for the Pomeron and Reggeon quark and gluon DPDFs.

At the highest expected EIC energy, we found that the partonic content of the Reggeon, taken from 
parametrizations of the pion in HERA studies, can be constrained at the EIC with uncertainties similar to those for the partonic content of the Pomeron. 
We observed that 
luminosity, choice of Monte Carlo sample and binning density have a marginal effect, while the crucial ingredient is the kinematic range in $\xi$ and $t$ covered by the pseudodata.

At the lowest EIC energy,
the Pomeron partonic content cannot be determined, because
the Reggeon becomes heavily dominant. 
Fixing the Pomeron contribution to 
that obtained in HERA data, we found that the DPDFs for the Reggeon can be determined with reasonable precision.

The largest limitation of our study comes from the model used to generate the pseudodata, which consists of  
only two 
components 
whose interference is neglected, following the studies at HERA and the model employed for generating the pseudodata~\cite{Chekanov:2009aa}.
While more sophisticated models 
could in principle be employed in future simulation
studies  (e.g., in~\cite{H1:1997vke} maximal and zero interference were compared), 
further insights into the real underlying dynamics of
colorless exchanges in DIS will have to wait 
until the arrival of real experimental data from the EIC.


\section*{Acknowledgments}

NA has received financial support from Xunta de Galicia (Centro singular de investigación de Galicia accreditation 2019-2022, ref. ED421G-2019/05), by European
Union ERDF, by the "María de Maeztu" Units of Excellence program MDM2016-
0692, and by the Spanish Research State Agency under project PID2020-119632GBI00. This work has been performed in the framework of the European Research
Council project ERC-2018-ADG-835105 YoctoLHC and the MSCA RISE 823947
"Heavy ion collisions: collectivity and precision in saturation physics" (HIEIC), and
has received funding from the European Union’s Horizon 2020 research and innovation programme under grant agreement No. 824093.
AMS is  supported by the U.S. Department of Energy grant No. DE-SC-0002145 and within the framework of the of the Saturated Glue (SURGE) Topical Theory Collaboration, as well as  in part by National Science Centre in Poland, grant 2019/33/B/ST2/02588.

\bibliography{mybib}

\begin{thebibliography}{33}
\expandafter\ifx\csname natexlab\endcsname\relax\def\natexlab#1{#1}\fi
\expandafter\ifx\csname bibnamefont\endcsname\relax
  \def\bibnamefont#1{#1}\fi
\expandafter\ifx\csname bibfnamefont\endcsname\relax
  \def\bibfnamefont#1{#1}\fi
\expandafter\ifx\csname citenamefont\endcsname\relax
  \def\citenamefont#1{#1}\fi
\expandafter\ifx\csname url\endcsname\relax
  \def\url#1{\texttt{#1}}\fi
\expandafter\ifx\csname urlprefix\endcsname\relax\def\urlprefix{URL }\fi
\providecommand{\bibinfo}[2]{#2}
\providecommand{\eprint}[2][]{\url{#2}}

\bibitem[{\citenamefont{Adloff et~al.}(1997{\natexlab{a}})}]{Adloff:1997sc}
\bibinfo{author}{\bibfnamefont{C.}~\bibnamefont{Adloff}} \bibnamefont{et~al.}
  (\bibinfo{collaboration}{H1}), \bibinfo{journal}{Z. Phys.}
  \textbf{\bibinfo{volume}{C76}}, \bibinfo{pages}{613}
  (\bibinfo{year}{1997}{\natexlab{a}}), \eprint{hep-ex/9708016}.

\bibitem[{\citenamefont{Breitweg et~al.}(1998)}]{Breitweg:1997aa}
\bibinfo{author}{\bibfnamefont{J.}~\bibnamefont{Breitweg}} \bibnamefont{et~al.}
  (\bibinfo{collaboration}{ZEUS}), \bibinfo{journal}{Eur. Phys. J.}
  \textbf{\bibinfo{volume}{C1}}, \bibinfo{pages}{81} (\bibinfo{year}{1998}),
  \eprint{hep-ex/9709021}.

\bibitem[{\citenamefont{Newman and Wing}(2014)}]{Newman:2013ada}
\bibinfo{author}{\bibfnamefont{P.}~\bibnamefont{Newman}} \bibnamefont{and}
  \bibinfo{author}{\bibfnamefont{M.}~\bibnamefont{Wing}},
  \bibinfo{journal}{Rev. Mod. Phys.} \textbf{\bibinfo{volume}{86}},
  \bibinfo{pages}{1037} (\bibinfo{year}{2014}), \eprint{1308.3368}.

\bibitem[{\citenamefont{Collins}(1998)}]{Collins:1997sr}
\bibinfo{author}{\bibfnamefont{J.~C.} \bibnamefont{Collins}},
  \bibinfo{journal}{Phys. Rev.} \textbf{\bibinfo{volume}{D57}},
  \bibinfo{pages}{3051} (\bibinfo{year}{1998}), \bibinfo{note}{[Erratum: Phys.
  Rev.D61,019902(2000)]}, \eprint{hep-ph/9709499}.

\bibitem[{\citenamefont{Berera and Soper}(1996)}]{Berera:1995fj}
\bibinfo{author}{\bibfnamefont{A.}~\bibnamefont{Berera}} \bibnamefont{and}
  \bibinfo{author}{\bibfnamefont{D.~E.} \bibnamefont{Soper}},
  \bibinfo{journal}{Phys. Rev.} \textbf{\bibinfo{volume}{D53}},
  \bibinfo{pages}{6162} (\bibinfo{year}{1996}), \eprint{hep-ph/9509239}.

\bibitem[{\citenamefont{Ingelman and Schlein}(1985)}]{Ingelman:1984ns}
\bibinfo{author}{\bibfnamefont{G.}~\bibnamefont{Ingelman}} \bibnamefont{and}
  \bibinfo{author}{\bibfnamefont{P.~E.} \bibnamefont{Schlein}},
  \bibinfo{journal}{Phys. Lett. B} \textbf{\bibinfo{volume}{152}},
  \bibinfo{pages}{256} (\bibinfo{year}{1985}).

\bibitem[{\citenamefont{Chekanov et~al.}(2005)}]{Chekanov:2005vv}
\bibinfo{author}{\bibfnamefont{S.}~\bibnamefont{Chekanov}} \bibnamefont{et~al.}
  (\bibinfo{collaboration}{ZEUS}), \bibinfo{journal}{Nucl. Phys.}
  \textbf{\bibinfo{volume}{B713}}, \bibinfo{pages}{3} (\bibinfo{year}{2005}),
  \eprint{hep-ex/0501060}.

\bibitem[{\citenamefont{Aktas et~al.}(2006{\natexlab{a}})}]{Aktas:2006hx}
\bibinfo{author}{\bibfnamefont{A.}~\bibnamefont{Aktas}} \bibnamefont{et~al.}
  (\bibinfo{collaboration}{H1 Collaboration}), \bibinfo{journal}{Eur.Phys.J.}
  \textbf{\bibinfo{volume}{C48}}, \bibinfo{pages}{749}
  (\bibinfo{year}{2006}{\natexlab{a}}), \eprint{hep-ex/0606003}.

\bibitem[{\citenamefont{Aktas et~al.}(2006{\natexlab{b}})}]{Aktas:2006hy}
\bibinfo{author}{\bibfnamefont{A.}~\bibnamefont{Aktas}} \bibnamefont{et~al.}
  (\bibinfo{collaboration}{H1 Collaboration}), \bibinfo{journal}{Eur.Phys.J.}
  \textbf{\bibinfo{volume}{C48}}, \bibinfo{pages}{715}
  (\bibinfo{year}{2006}{\natexlab{b}}), \eprint{hep-ex/0606004}.

\bibitem[{\citenamefont{Chekanov et~al.}(2009)}]{Chekanov:2008fh}
\bibinfo{author}{\bibfnamefont{S.}~\bibnamefont{Chekanov}} \bibnamefont{et~al.}
  (\bibinfo{collaboration}{ZEUS Collaboration}), \bibinfo{journal}{Nucl.Phys.}
  \textbf{\bibinfo{volume}{B816}}, \bibinfo{pages}{1} (\bibinfo{year}{2009}),
  \eprint{0812.2003}.

\bibitem[{\citenamefont{Chekanov et~al.}(2010)}]{Chekanov:2009aa}
\bibinfo{author}{\bibfnamefont{S.}~\bibnamefont{Chekanov}} \bibnamefont{et~al.}
  (\bibinfo{collaboration}{ZEUS Collaboration}), \bibinfo{journal}{Nucl.Phys.}
  \textbf{\bibinfo{volume}{B831}}, \bibinfo{pages}{1} (\bibinfo{year}{2010}),
  \eprint{0911.4119}.

\bibitem[{\citenamefont{Aaron et~al.}(2011)}]{Aaron:2010aa}
\bibinfo{author}{\bibfnamefont{F.~D.} \bibnamefont{Aaron}}
  \bibnamefont{et~al.}, \bibinfo{journal}{Eur. Phys. J. C}
  \textbf{\bibinfo{volume}{71}}, \bibinfo{pages}{1578} (\bibinfo{year}{2011}),
  \eprint{1010.1476}.

\bibitem[{\citenamefont{Aaron et~al.}(2012{\natexlab{a}})}]{Aaron:2012ad}
\bibinfo{author}{\bibfnamefont{F.}~\bibnamefont{Aaron}} \bibnamefont{et~al.}
  (\bibinfo{collaboration}{H1 Collaboration}), \bibinfo{journal}{Eur.Phys.J.}
  \textbf{\bibinfo{volume}{C72}}, \bibinfo{pages}{2074}
  (\bibinfo{year}{2012}{\natexlab{a}}), \eprint{1203.4495}.

\bibitem[{\citenamefont{Gluck et~al.}(1992)\citenamefont{Gluck, Reya, and
  Vogt}}]{Gluck:1991ey}
\bibinfo{author}{\bibfnamefont{M.}~\bibnamefont{Gluck}},
  \bibinfo{author}{\bibfnamefont{E.}~\bibnamefont{Reya}}, \bibnamefont{and}
  \bibinfo{author}{\bibfnamefont{A.}~\bibnamefont{Vogt}}, \bibinfo{journal}{Z.
  Phys.} \textbf{\bibinfo{volume}{C53}}, \bibinfo{pages}{651}
  (\bibinfo{year}{1992}).

\bibitem[{\citenamefont{Accardi et~al.}(2016)}]{Accardi:2012qut}
\bibinfo{author}{\bibfnamefont{A.}~\bibnamefont{Accardi}} \bibnamefont{et~al.},
  \bibinfo{journal}{Eur. Phys. J.} \textbf{\bibinfo{volume}{A52}},
  \bibinfo{pages}{268} (\bibinfo{year}{2016}), \eprint{1212.1701}.

\bibitem[{\citenamefont{Abdul~Khalek et~al.}(2022)}]{AbdulKhalek:2021gbh}
\bibinfo{author}{\bibfnamefont{R.}~\bibnamefont{Abdul~Khalek}}
  \bibnamefont{et~al.}, \bibinfo{journal}{Nucl. Phys. A}
  \textbf{\bibinfo{volume}{1026}}, \bibinfo{pages}{122447}
  (\bibinfo{year}{2022}), \eprint{2103.05419}.

\bibitem[{\citenamefont{Gribov and
  Lipatov}(1972{\natexlab{a}})}]{Gribov:1972ri}
\bibinfo{author}{\bibfnamefont{V.~N.} \bibnamefont{Gribov}} \bibnamefont{and}
  \bibinfo{author}{\bibfnamefont{L.~N.} \bibnamefont{Lipatov}},
  \bibinfo{journal}{Sov. J. Nucl. Phys.} \textbf{\bibinfo{volume}{15}},
  \bibinfo{pages}{438} (\bibinfo{year}{1972}{\natexlab{a}}),
  \bibinfo{note}{[Yad. Fiz.15,781(1972)]}.

\bibitem[{\citenamefont{Gribov and
  Lipatov}(1972{\natexlab{b}})}]{Gribov:1972rt}
\bibinfo{author}{\bibfnamefont{V.~N.} \bibnamefont{Gribov}} \bibnamefont{and}
  \bibinfo{author}{\bibfnamefont{L.~N.} \bibnamefont{Lipatov}},
  \bibinfo{journal}{Sov. J. Nucl. Phys.} \textbf{\bibinfo{volume}{15}},
  \bibinfo{pages}{675} (\bibinfo{year}{1972}{\natexlab{b}}),
  \bibinfo{note}{[Yad. Fiz.15,1218(1972)]}.

\bibitem[{\citenamefont{Dokshitzer}(1977)}]{Dokshitzer:1977sg}
\bibinfo{author}{\bibfnamefont{Y.~L.} \bibnamefont{Dokshitzer}},
  \bibinfo{journal}{Sov. Phys. JETP} \textbf{\bibinfo{volume}{46}},
  \bibinfo{pages}{641} (\bibinfo{year}{1977}), \bibinfo{note}{[Zh. Eksp. Teor.
  Fiz.73,1216(1977)]}.

\bibitem[{\citenamefont{Altarelli and Parisi}(1977)}]{Altarelli:1977zs}
\bibinfo{author}{\bibfnamefont{G.}~\bibnamefont{Altarelli}} \bibnamefont{and}
  \bibinfo{author}{\bibfnamefont{G.}~\bibnamefont{Parisi}},
  \bibinfo{journal}{Nucl. Phys.} \textbf{\bibinfo{volume}{B126}},
  \bibinfo{pages}{298} (\bibinfo{year}{1977}).

\bibitem[{\citenamefont{Trentadue and Veneziano}(1994)}]{Trentadue:1993ka}
\bibinfo{author}{\bibfnamefont{L.}~\bibnamefont{Trentadue}} \bibnamefont{and}
  \bibinfo{author}{\bibfnamefont{G.}~\bibnamefont{Veneziano}},
  \bibinfo{journal}{Phys. Lett.} \textbf{\bibinfo{volume}{B323}},
  \bibinfo{pages}{201} (\bibinfo{year}{1994}).

\bibitem[{\citenamefont{Adkins et~al.}(2022)}]{Adkins:2022jfp}
\bibinfo{author}{\bibfnamefont{J.~K.} \bibnamefont{Adkins}}
  \bibnamefont{et~al.} (\bibinfo{year}{2022}), \eprint{2209.02580}.

\bibitem[{\citenamefont{Adam et~al.}(2022)}]{ATHENA:2022hxb}
\bibinfo{author}{\bibfnamefont{J.}~\bibnamefont{Adam}} \bibnamefont{et~al.}
  (\bibinfo{collaboration}{ATHENA}), \bibinfo{journal}{JINST}
  \textbf{\bibinfo{volume}{17}}, \bibinfo{pages}{P10019}
  (\bibinfo{year}{2022}), \eprint{2210.09048}.

\bibitem[{epi()}]{epic}
\emph{\bibinfo{title}{The epic collaboration}},
  \bibinfo{howpublished}{https://www.bnl.gov/eic/epic.php}.

\bibitem[{\citenamefont{Armesto et~al.}(2022)\citenamefont{Armesto, Newman,
  Slominski, and Stasto}}]{Armesto:2021fws}
\bibinfo{author}{\bibfnamefont{N.}~\bibnamefont{Armesto}},
  \bibinfo{author}{\bibfnamefont{P.~R.} \bibnamefont{Newman}},
  \bibinfo{author}{\bibfnamefont{W.}~\bibnamefont{Slominski}},
  \bibnamefont{and} \bibinfo{author}{\bibfnamefont{A.~M.}
  \bibnamefont{Stasto}}, \bibinfo{journal}{Phys. Rev. D}
  \textbf{\bibinfo{volume}{105}}, \bibinfo{pages}{074006}
  (\bibinfo{year}{2022}), \eprint{2112.06839}.

\bibitem[{\citenamefont{Cerci et~al.}(2023)\citenamefont{Cerci, Demiroglu,
  Deshpande, Newman, Schmookler, Sunar~Cerci, and Wichmann}}]{Cerci:2023uhu}
\bibinfo{author}{\bibfnamefont{S.}~\bibnamefont{Cerci}},
  \bibinfo{author}{\bibfnamefont{Z.~S.} \bibnamefont{Demiroglu}},
  \bibinfo{author}{\bibfnamefont{A.}~\bibnamefont{Deshpande}},
  \bibinfo{author}{\bibfnamefont{P.~R.} \bibnamefont{Newman}},
  \bibinfo{author}{\bibfnamefont{B.}~\bibnamefont{Schmookler}},
  \bibinfo{author}{\bibfnamefont{D.}~\bibnamefont{Sunar~Cerci}},
  \bibnamefont{and} \bibinfo{author}{\bibfnamefont{K.}~\bibnamefont{Wichmann}},
  \bibinfo{journal}{Eur. Phys. J. C} \textbf{\bibinfo{volume}{83}},
  \bibinfo{pages}{1011} (\bibinfo{year}{2023}), \eprint{2307.01183}.

\bibitem[{\citenamefont{Armesto et~al.}(2024)\citenamefont{Armesto, Cridge,
  Giuli, Harland-Lang, Newman, Schmookler, Thorne, and
  Wichmann}}]{Armesto:2023hnw}
\bibinfo{author}{\bibfnamefont{N.}~\bibnamefont{Armesto}},
  \bibinfo{author}{\bibfnamefont{T.}~\bibnamefont{Cridge}},
  \bibinfo{author}{\bibfnamefont{F.}~\bibnamefont{Giuli}},
  \bibinfo{author}{\bibfnamefont{L.}~\bibnamefont{Harland-Lang}},
  \bibinfo{author}{\bibfnamefont{P.}~\bibnamefont{Newman}},
  \bibinfo{author}{\bibfnamefont{B.}~\bibnamefont{Schmookler}},
  \bibinfo{author}{\bibfnamefont{R.}~\bibnamefont{Thorne}}, \bibnamefont{and}
  \bibinfo{author}{\bibfnamefont{K.}~\bibnamefont{Wichmann}},
  \bibinfo{journal}{Phys. Rev. D} \textbf{\bibinfo{volume}{109}},
  \bibinfo{pages}{054019} (\bibinfo{year}{2024}), \eprint{2309.11269}.

\bibitem[{\citenamefont{Aaron et~al.}(2012{\natexlab{b}})}]{H1:2012xlc}
\bibinfo{author}{\bibfnamefont{F.~D.} \bibnamefont{Aaron}} \bibnamefont{et~al.}
  (\bibinfo{collaboration}{H1, ZEUS}), \bibinfo{journal}{Eur. Phys. J. C}
  \textbf{\bibinfo{volume}{72}}, \bibinfo{pages}{2175}
  (\bibinfo{year}{2012}{\natexlab{b}}), \eprint{1207.4864}.

\bibitem[{\citenamefont{Adloff et~al.}(1997{\natexlab{b}})}]{H1:1997vke}
\bibinfo{author}{\bibfnamefont{C.}~\bibnamefont{Adloff}} \bibnamefont{et~al.}
  (\bibinfo{collaboration}{H1}), \bibinfo{journal}{Z. Phys. C}
  \textbf{\bibinfo{volume}{74}}, \bibinfo{pages}{221}
  (\bibinfo{year}{1997}{\natexlab{b}}), \eprint{hep-ex/9702003}.

\bibitem[{\citenamefont{Armesto et~al.}(2019)\citenamefont{Armesto, Newman,
  S\l{}omi\'nski, and Sta\'sto}}]{Armesto:2019gxy}
\bibinfo{author}{\bibfnamefont{N.}~\bibnamefont{Armesto}},
  \bibinfo{author}{\bibfnamefont{P.~R.} \bibnamefont{Newman}},
  \bibinfo{author}{\bibfnamefont{W.}~\bibnamefont{S\l{}omi\'nski}},
  \bibnamefont{and} \bibinfo{author}{\bibfnamefont{A.~M.}
  \bibnamefont{Sta\'sto}}, \bibinfo{journal}{Phys. Rev. D}
  \textbf{\bibinfo{volume}{100}}, \bibinfo{pages}{074022}
  (\bibinfo{year}{2019}), \eprint{1901.09076}.

\bibitem[{\citenamefont{Collins and Tung}(1986)}]{Collins:1986mp}
\bibinfo{author}{\bibfnamefont{J.~C.} \bibnamefont{Collins}} \bibnamefont{and}
  \bibinfo{author}{\bibfnamefont{W.-K.} \bibnamefont{Tung}},
  \bibinfo{journal}{Nucl. Phys.} \textbf{\bibinfo{volume}{B278}},
  \bibinfo{pages}{934} (\bibinfo{year}{1986}).

\bibitem[{\citenamefont{Thorne and Tung}(2008)}]{Thorne:2008xf}
\bibinfo{author}{\bibfnamefont{R.~S.} \bibnamefont{Thorne}} \bibnamefont{and}
  \bibinfo{author}{\bibfnamefont{W.~K.} \bibnamefont{Tung}}
  (\bibinfo{year}{2008}), \eprint{0809.0714}.

\bibitem[{\citenamefont{Pumplin et~al.}(2001)\citenamefont{Pumplin, Stump,
  Brock, Casey, Huston, Kalk, Lai, and Tung}}]{Pumplin:2001ct}
\bibinfo{author}{\bibfnamefont{J.}~\bibnamefont{Pumplin}},
  \bibinfo{author}{\bibfnamefont{D.}~\bibnamefont{Stump}},
  \bibinfo{author}{\bibfnamefont{R.}~\bibnamefont{Brock}},
  \bibinfo{author}{\bibfnamefont{D.}~\bibnamefont{Casey}},
  \bibinfo{author}{\bibfnamefont{J.}~\bibnamefont{Huston}},
  \bibinfo{author}{\bibfnamefont{J.}~\bibnamefont{Kalk}},
  \bibinfo{author}{\bibfnamefont{H.~L.} \bibnamefont{Lai}}, \bibnamefont{and}
  \bibinfo{author}{\bibfnamefont{W.~K.} \bibnamefont{Tung}},
  \bibinfo{journal}{Phys. Rev. D} \textbf{\bibinfo{volume}{65}},
  \bibinfo{pages}{014013} (\bibinfo{year}{2001}), \eprint{hep-ph/0101032}.

\end{thebibliography}

\end{document}